\documentclass[useAMS,usenatbib]{mn2e}
\usepackage{amsmath}
\usepackage{graphics}
\usepackage{graphicx}
\usepackage{color}
\usepackage{natbib}
\newcommand{\apj}{ApJ}

\newcommand{\mnras}{MNRAS}

\newcommand{\be}{\begin{equation}}
\newcommand{\ee}{\end{equation}}
\newcommand{\ba}{\begin{eqnarray}}
\newcommand{\ea}{\end{eqnarray}}
\newcommand{\brr}{\begin{array}}

\newcommand{\err}{\end{array}}
\newcommand{\bc}{\begin{center}}
\newcommand{\ec}{\end{center}}

\newcommand{\ie}{\it i.e.}
\newcommand{\Bpos}{\mbox{$B_{e^{+}}$}}
\newcommand{\ebar}{\mbox{$e^{+}$}}
\newcommand{\pbar}{\mbox{$\bar{\rm p}$}}
\newcommand{\Bpbar}{\mbox{$B_{\bar{\rm p}}$}}
\newcommand{\Gpbar}{\mbox{$G_{\bar{\rm p}}$}}
\newcommand{\lD}{\mbox{$\lambda_{\rm D}$}}
\newcommand{\aaa}{\hspace{0.1cm}}
\begin{document}
\title[{The Cosmic Ray Signature of Dark Matter Caustics}]
{The Cosmic Ray Signature of Dark Matter Caustics}
\author[{Roya Mohayaee \& Pierre Salati}]
{Roya Mohayaee$^1$, Pierre Salati$^2$\\
$^1$Institut d'Astrophysique de Paris (IAP), CNRS, UPMC, 
98 bis boulevard Arago, France\\
$^2$LAPTH, 9 Chemin de Bellevue, BP110, F-74941 Annecy-le-Vieux cedex, France\\
emails: roya@iap.fr ; salati@lapp.in2p3.fr
}
\maketitle
\begin{abstract}
Gravitational collapse of dark matter, merger of dark matter haloes and tidal
disruption of satellites are among processes which lead to the formation of fine
and dense dark matter shells, also known as dark matter caustics.
The putative weakly interacting species which may form the dark matter are
expected to strongly annihilate in these dense regions of the Milky Way halo and
generate in particular antiprotons and positrons. We derive the flux of these rare
antimatter particles at the Earth and show that it depends significantly on the
cut-off radius of the dark matter distribution at the galactic centre.
Boost factors of $\sim 30$ are found with respect to a smooth NFW profile
for high-energy antiprotons and low-energy positrons if this cut-off radius
is taken to be 300 pc -- a somewhat extreme value though.
This yields a detectable antiproton signal around hundreds of Gev
in models where the annihilation cross section today is enhanced by non--perturbative
effects as in the generic case of a heavy Wino.
However, dark matter caustics cannot provide a better explanation for the HEAT excess
reported above $\sim$ 10 GeV than a smooth NFW or isothermal cored distribution.
%

\noindent
{\small\sf preprint no: LAPTH-1235/08}
\end{abstract}

\maketitle
\hspace{.2in}

%
\section{Introduction}
\label{sec:introduction}

The halo of our galaxy is believed to have formed from the gravitational
collapse of dark matter (DM) and a continuous merger with other haloes. 
The formation of haloes, described by the Jeans-Vlasov-Poisson equation,
proceeds with the formation of
dark matter {\it shells}, also referred to
as dark matter {\it caustics}
\citep{sikivieipser,sikivietkachev,tremaine,alardcolombi,natarajan}. 
As the density peaks collapse under
self-gravity, at the surfaces of (formally) infinite density 
where dark matter streams meet, caustics emerge. 
The infalling satellites are also distrupted by a mechanism similar to
the formation of the {\it primordial} caustics. As a satellite
moves inside the potential well of its host halo, a tidal tail 
forms, and around
the apapsis of the orbit high-density shells or caustics emerge
[see {\it e.g.} \citet{hayashi}]. Since each infalling satellite also has his own caustics, the
hierarchical formation of structures yields a hierarchy of caustics.
Observational examples for the caustics 
are the shell galaxies, where shells of stars form during the
merger of galaxies, by a mechanism very similar to the formation
of dark matter caustics
\citep{malincarter,carterallen,hernquistquinn1,hernquistquinn2}.

Once formed, they
cannot be destroyed~: caustics are permanent structures in real and in
phase space. However, their
over-density with respect to the background 
density can diminish with time 
due to continuous increase in the number of streams.
Therefore, it is likely that we are surrounded by a large
number of primordial shells and fossil shells of disrupted satellites. 
The density of the shells can exceed that of the
diffuse background by an amount which depends on the age of the galaxy
and the number of times the
satellite has wrapped around our galaxy. 
Due to their vicinity and
over-density, these shells can be important 
for dark matter search experiments. 

The nature of dark matter remains a
mystery. Supersymmetry and
extra-dimension extensions of the standard electroweak model provide
a natural candidate in the form of a weakly interacting and massive particle
(hereafter WIMP). These species should fill up the galactic halo.
If dark matter consists of WIMPS then they are 
expected to strongly annihilate in the dense regions of our halo ({\it
i.e.} in the caustics) and generate in particular
gamma-rays and charged cosmic rays such as positrons and antiprotons.
Indeed, due to their very short diffusion length, positrons can be
excellent tracers of nearby caustics.
Although the implication of substructures, or surviving satellites, in
boosting the cosmic-ray signal has been studied extensively 
[see {\it e.g.} \citet{Lavalle:2006vb,Lavalle_Maurin}], the signature of
dark matter caustics on cosmic-rays has been rarely considered.  

Here, we propose to study the boost in positron and antiproton signals due to
annihilation in dark matter caustics. The ideal way to pursue this study
would be to use a
very high-resolution simulation. However, even the latest
simulations have not yet achieved a high enough 
resolution to identify the caustics in
a typical galaxy, although impressive progresses are being made
\citep{vogelsberger}. 
Analytic models for the formation of caustics are mainly based on the
self-similar secondary infall \citep{fillmoregoldreich,bertschinger}.
We use a generalized version of the secondary infall model 
which takes into account the finite velocity dispersion of DM
\citep{mohayaeeshandarin}. Although secondary infall model
has serious limitations because of the assumption of spherical symmetry and 
smooth accretion and because it ignores the hierarchical scenario of structure
formation, it has so far provided a paradigm for the study of
dark matter haloes [{\it e.g.} see \citet{ascasibar}].

In Section~2, we review and generalize the secondary infall model for dark matter
with finite velocity dispersion and give an analytic fit for the density profile 
which includes the caustics for a typical galaxy like our own.
The cosmic ray signature of dark matter caustics is studied in Section~3
for large antiproton and positron horizon radii.
In Section~4, the possibility of detecting a nearby caustic through
its short range positron signal is investigated.
We finally conclude in Section~5.

%
\begin{figure*}
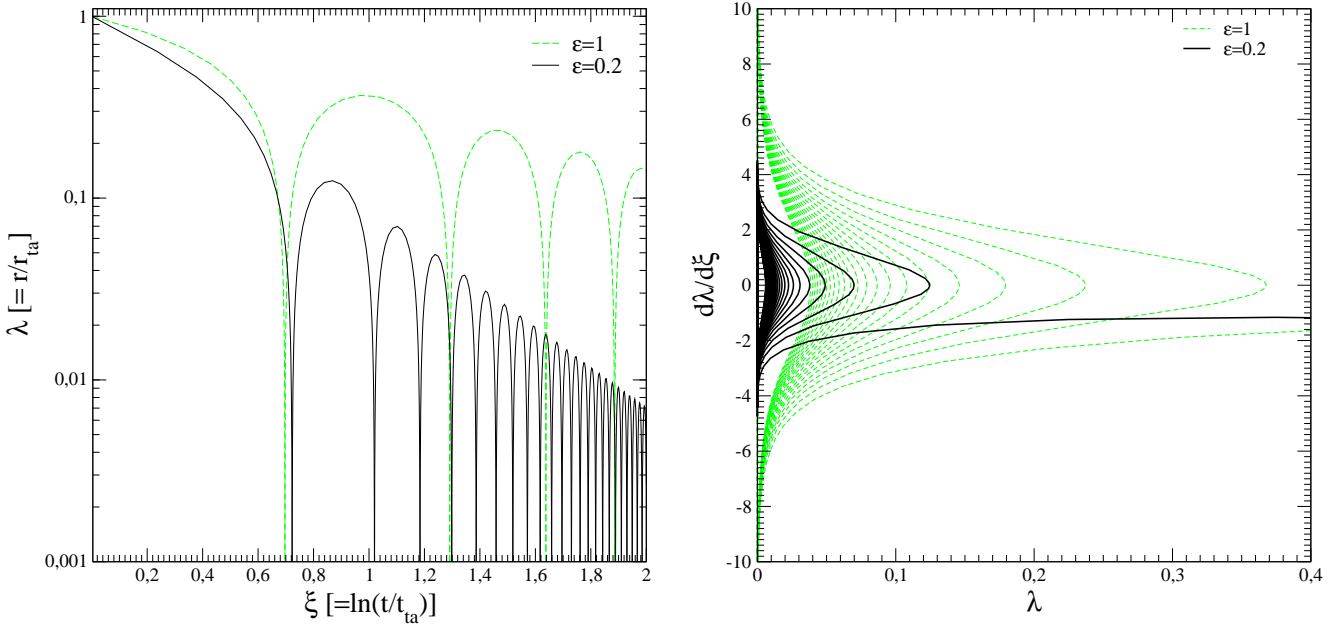

\centering{
\includegraphics[width=\columnwidth]{trajectory-pruned}
{\hspace{0.2cm}}
\includegraphics[width=\columnwidth]{phase-space-pruned}}
\caption{
Particle trajectories in real and phase spaces are shown for two values
of the parameter $\epsilon$ where $\lambda$ is the
non-dimensional radius and $\xi$ is the non-dimensional time.
The curves for $\epsilon=1$ are featured in order to highlight
the phase transition that occurs at $\epsilon=2/3$. The physical origin for
that transition is that for $\epsilon<2/3$, at late times, the mass within
a given radius $r$ is dominated by the contribution from particles from outside
of $r$. In contrast for $\epsilon>2/3$, the mass is dominated by particles within
$r$ (see \citep{fillmoregoldreich} for further details).
}
\label{fig:trajectory}
\end{figure*}
%
%
\begin{figure*}
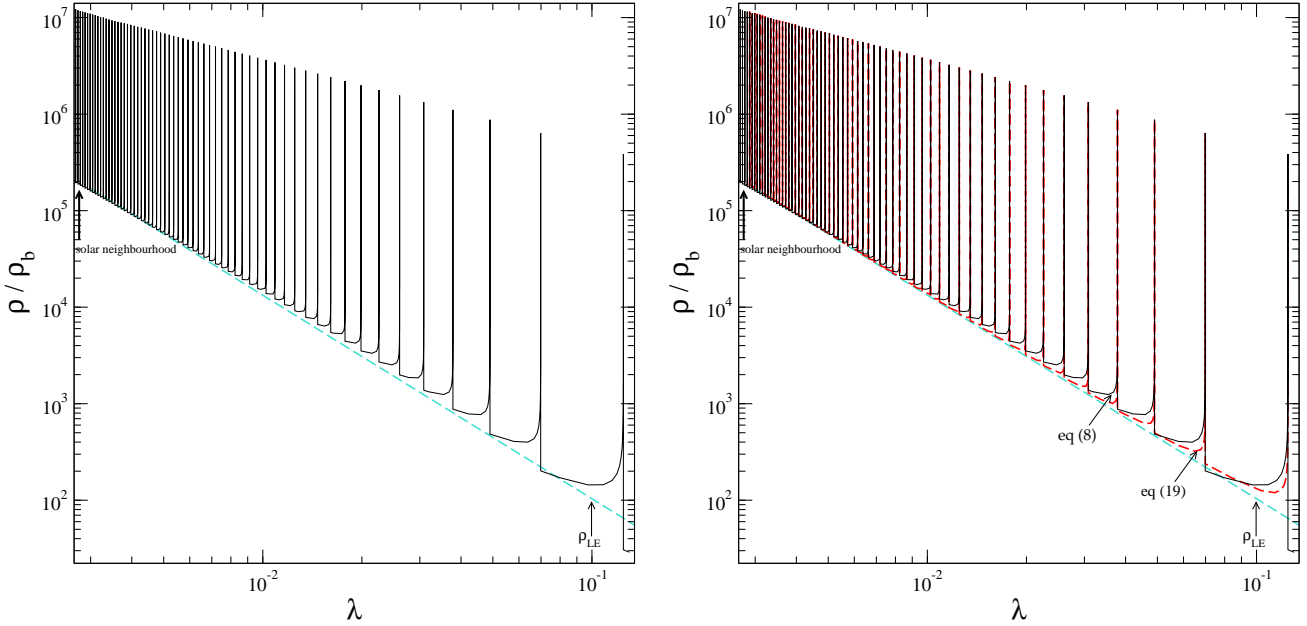

\centering{
\includegraphics[width=\columnwidth]{density-MW-pruned}
{\hspace{0.2cm}}
\includegraphics[width=\columnwidth]{density-numeric-analytic-pruned}}
\caption{
The left panel features the total numerically-obtained density
for $\epsilon=0.2$ [solid black curve, using expression~(\ref{eq:total density})].
This solution is compared in the right panel to our analytic fit [dashed red
curve, using expression~(\ref{eq:rho_total})].
}
\label{fig:total density}
\end{figure*}
%

%
\section{The secondary infall or self-similar accretion model}
\label{sec:self_similar}

%
\begin{figure*}
\includegraphics[width=\columnwidth]{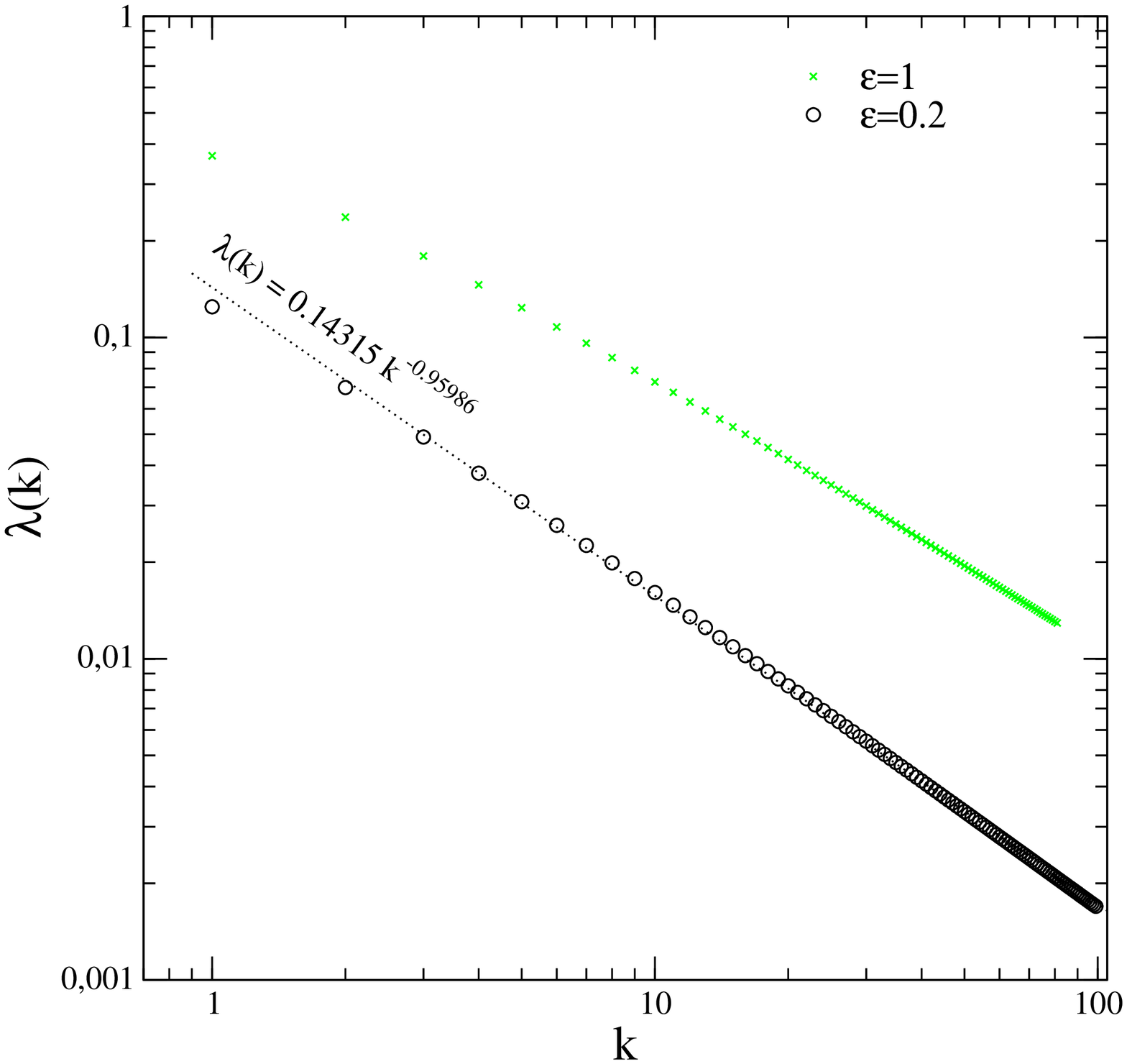}
\includegraphics[width=\columnwidth]{xi-k}
\includegraphics[width=\columnwidth]{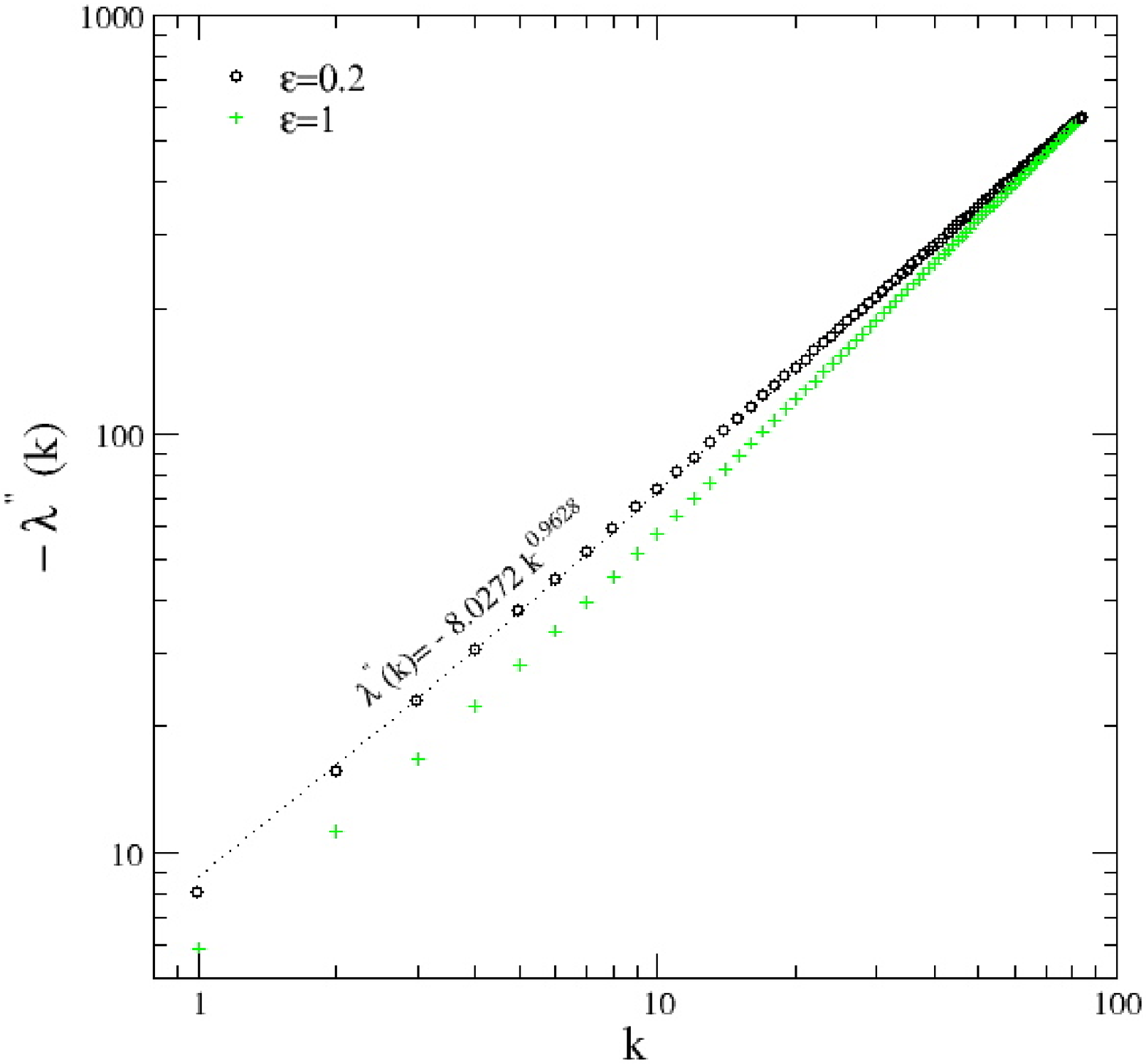}
\includegraphics[width=\columnwidth]{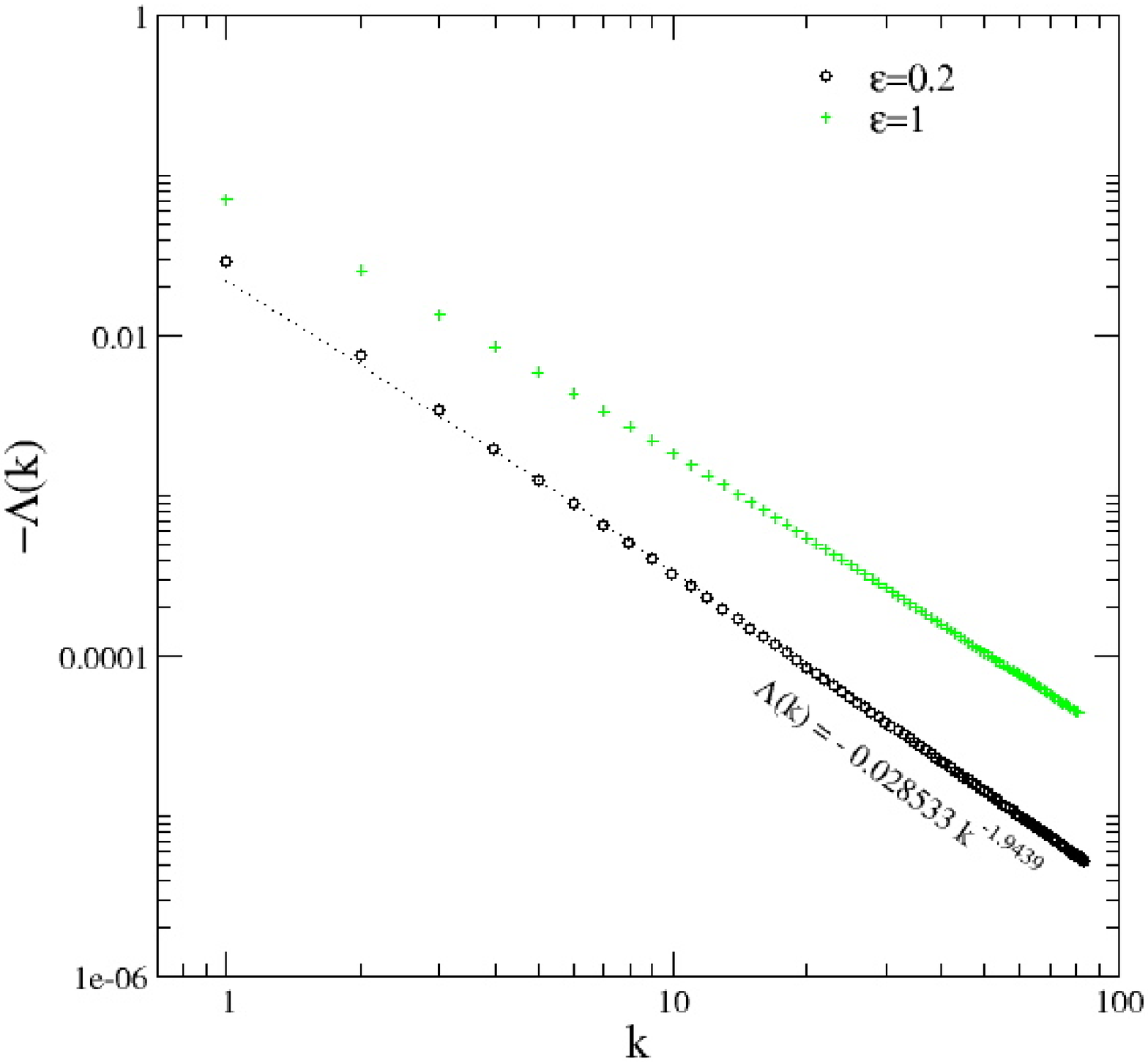}
\caption{
The dotted lines are the fittings
[presented in expressions~(\ref{eq:lambda fit}) to (\ref{eq:Lambda fit})]
to numerical results from integration of the differential equation~(\ref{eq:equation-motion}).
The black open circles and the green crosses correspond respectively to $\epsilon = 0.2$ and
$\epsilon = 1$.
}
\label{fig:fitting to caustic parameters}
\end{figure*}
%

Analytic evaluation of the halo
density profile, and the prediction of the existence of caustics inside these
structures, started with the works of
\citet{Gott} and \citet{Gunn}. With the main objective of
explaining the flattening of the rotation curves of the galaxies,
they considered the formation of a dark matter halo
from the infall of matter onto an already-formed galaxy (or in
later works onto a spherical overdense region).
In an Einstein-de Sitter Universe a spherical over-density
expands and then turns around to collapse.
After collapse and at late times, the fluid motion becomes
self-similar~: its form
remains unchanged when length are re-scaled in terms of $r_{ta}$, the radius of the 
shell which is currently turning around and collapsing.
Physically self-similarity arises because gravity is
scale-free and because mass shells outside the initial over-density are
also bound and turnaround at successively later times.
Self-similar solutions give power-law density profiles
whose exact scaling properties depend on the initial condition, the central boundary
conditions and on whether the fluid is collisionless or collisional
\citep{fillmoregoldreich,bertschinger}. The global density profile
is a power-law which for suitable choices of initial conditions provides an explanation of the
flattening of the rotation curves of the galaxies. A density profile of
$r^{-2}$ results for the galactic-scale fluctuations.
However, the
density profile is convolved with many spikes
({\it i.e.} caustics) of {\it formally} infinite densities which 
are rendered finite mainly by the velocity
dispersion of dark matter. 

In this paper, although we consider an Einstein-de Sitter Universe, we
do expect our results to give reasonable approximation for a $\Lambda$CDM
Universe as well. 
 The role of dark energy
becomes significant at rather small redshifts ($\sim 0.2$) which
we expect to be well after the formation of the typical dark matter haloes we
consider here. Furthermore, once a particle turns around and collapses, it
separates from the background expansion and its subsequent motion should not
be affected by the $\Lambda$ term. 

The effect of angular momentum is not considered here and we do not
expect it to have a significant effect far from the {\it core}
of dark matter haloes. Indeed detailed
studies show that angular momentum will not alter the results of 
the self-similar model (e.g. see \citep{nusser2001}).

In this model, the initial perturbation is assumed to be scale free, {\it i.e.}
\be
\delta = {\displaystyle \frac{\delta M}{M}} \propto
M^{\displaystyle - \epsilon}
\qquad {\rm at} \qquad
t = t_i \;\; .
\label{eq:first-epsilon}
\ee
Such scale-free perturbations result naturally in cosmological models with
scale-free initial spectra. For a power-law spectrum $P(k)\sim k^n$, the mass
variance $\sigma(M)$ scales as $M^{-(n+3)/6}$,
which yields 
\be
\epsilon = {n+3\over 6} \;\; .
\ee

As more and more mass shells turn around and collapse, the mass of halo grows 
at a rate of $t^{2/3\epsilon}$.
The problem exhibits similarity solution {\it asymptotically } 
and can be solved numerically. The numerical solutions confirm the
theoretical results which have been obtained assuming 
adiabatic invariance \citep{fillmoregoldreich}.
Asymptotically, the density profile reaches the power-law~:
\be
\rho = \left\{
\begin{array}{lcl}
r^{\displaystyle -2} & \qquad ; \qquad &
\epsilon \le \frac{2}{3} \\
\\
r^{\displaystyle -9 \epsilon / (1 + 3 \epsilon)} & \qquad ; \qquad &
\epsilon \ge \frac{2}{3} \\
\end{array}
\right.
\label{eq:smooth-density-profile-epsilon}
\ee
For galactic-scale fluctuations ($n \sim 1.2$ as
given by the galaxy-galaxy correlation function)
one has $\epsilon\sim 0.2-0.3$. Here, we set $\epsilon = 0.2$, also for the purpose
of comparing a few of our results with \citet{fillmoregoldreich}.
The evolution equation, namely Newton's equation of motion written in
self-similar variables, is \footnote{This is equation (23) of 
\citet{fillmoregoldreich}, rewritten in variables~(\ref{eq:Bvariables})
[also used in \citet{bertschinger}].}
\ba
{\displaystyle \frac{d^{2}\lambda}{d\xi^{2}}} & \!\!\! + \!\!\! &
\left( {\displaystyle \frac{1}{3}} + {\displaystyle \frac{4}{9 \epsilon}} \right)
{\displaystyle \frac{d\lambda}{d\xi}}  \, + \,
\left( {\displaystyle \frac{2}{3}} + {\displaystyle \frac{2}{9 \epsilon}} \right)
\left( {\displaystyle \frac{2}{9 \epsilon}} - {\displaystyle \frac{1}{3}} \right)
\lambda \nonumber \\
& \!\! = \!\! &
- \, {\displaystyle \frac{2}{9 \lambda^{2}}} \;
\exp \left( {\displaystyle \frac{2}{3 \epsilon}} (\epsilon - 1) \xi
\right) \;
M(\lambda) \;\; ,
\label{eq:equation-motion}
\ea
where
\be
\lambda = {\displaystyle \frac{r}{r_{ta}}}
\qquad {\rm and} \qquad
\xi = {\rm ln} \left( {\displaystyle \frac{t}{t_{\rm ta}}} \right) \;\; .
\label{eq:Bvariables}
\ee
The dimensionless radius $\lambda$ is given in terms of the physical
radius $r$ and present turnaround radius $r_{\rm ta}$
and $\xi$ is the dimensionless time
given in terms of the turnaround time $t_{\rm ta}$ for each particle.
Equation~(\ref{eq:equation-motion}) has been solved with the 
initial condition given at $\xi = 0$ (corresponding to $t = t_{\rm ta}$)
\be
\lambda = 1
\qquad {\rm and} \qquad
{\displaystyle \frac{d\lambda}{d\xi}} = - P \;\; ,
\label{initialcondition}
\ee
with a prior knowledge of the mass $M(\lambda)$ and where
\be
P={2\over 3}\left(1+{1\over 3\epsilon}\right)\;\;.
\ee
The solution to (\ref{eq:equation-motion}) can been obtained iteratively~: following an initial guess for
$M(\lambda)$, equation~(\ref{eq:equation-motion}) is integrated, then
$M(\lambda)$ is evaluated, and the procedure is continued until the desired
level of convergence is achieved.
Here, we take a simpler approach. At small values
of $\lambda$ ($\lambda \ll 1$), mass becomes a 
power-law $M(\lambda)\approx \lambda$ (Fillmore \& Goldreich 1984). 
We thus fit $M(\lambda)$ for $\epsilon=0.2$ by the following profile
\be
{\displaystyle \frac{M(\lambda)}{M_{\rm ta}}} =
{\lambda \over 1 + \lambda} \;\; ,
\label{eq:massoverturnaroundmass}
\ee
where $M_{ta}=(3\pi/4)^2$. The solutions are shown in Fig.~\ref{fig:trajectory}
for two choices of $\epsilon$. 
The approximation (\ref{eq:massoverturnaroundmass}) generates relatively small errors.
A notable discrepancy between the results obtained through approximation
(\ref{eq:massoverturnaroundmass}) and by iteration
appears only at the relatively large values 
of $\lambda\sim 1$. However, for all
the caustics under consideration the value of $\lambda$ is far less than one.
The largest value of $\lambda$ for 
the first caustic is at $\lambda\approx 0.36$ for $\epsilon=1$ and at
$\lambda\approx 0.12$ for $\epsilon=0.2$ (see the table in Fig. A1 
of \citet{mohayaeeshandarin} for further details).

The density profile is given by
\be
{\rho\over\rho_b} = {\pi^2 \over 8 \lambda^2} \, \sum_j \,
(-1)^j \, {\rm exp}\left(-{2\over 3\epsilon} \, \xi_j\right)
\left({d\lambda\over d\xi}\right)_j^{-1}\;\;,
\label{eq:total density}
\ee
where $\rho_b$ is the cosmological background density which we take
to be $1.6~\times~10^{-6}$ GeV cm$^{-3}$ and 
$t_{\rm ta}$ is the turnaround time of the particle that is
at the $j$th point where $\lambda=\lambda(\xi)$ 
[see \citet{bertschinger} for further explanation].

The density~(\ref{eq:total density}) is evaluated numerically and plotted in
Fig.~\ref{fig:total density}. 
Evidently, the density profile is convolved 
with many spikes, namely caustics, where the density formally diverges.
However, there is a natural cut-off to the density of the caustics due to the
finite velocity dispersion of dark matter \citep{mohayaeeshandarin}. 
For a generic value of $\epsilon$, 
the thickness of the caustic shell may be expressed as
\be
\Delta \lambda_k = {(3\pi)^{2/3}\over 4}\,\left({4\over
3\epsilon+1}\right)^{1/3} 
\, e^{(P-1/3)\xi_k} \, \Lambda_k
\; {t \, \sigma(t) \over r_{\rm ta}} \;\; .
\label{eq:caustic thickness}
\ee
The maximum density at the caustic positions is given by
\be
\rho_{k,{\sf max}}={G_k\over\sqrt{|\Delta\lambda_k|}} \, \rho_b \;\; ,
\label{eq:maximum density}
\ee
where
\be
G_k = {9\, P \, \pi^2\over 32 \sqrt{-2\lambda^{\prime\prime}_k}}\left({4\over
3\epsilon+1}\right)^{5/6}
\, {e^{(6-9\,P)\xi_k/3} \over \lambda_k^2}\;\;,
\ee
and $t$ is the age of the Universe,
$\sigma$ is the present-day velocity dispersion of dark matter particles which is that
at  decoupling re-scaled with the expansion factor
and $\xi_k={\rm ln}(t/t_{ta})$ is the dimensionless time given in terms of the turnaround
time $t_{\rm ta}$ of the particle that is now in caustic $k$.
The first ten values of these parameters are listed in
Table~\ref{table:epsilon0o2}, assuming a velocity dispersion
$\sigma$ of $0.03$ cm s$^{-1}$.

The density profile of caustics is simply given by
\be
\rho_ k = 
\left\{
\begin{array}{ll}
\rho_{k,{\sf max}} &\quad \, 
{\lambda_k - |\Delta\lambda_k| \le \lambda \le \lambda_k} \;\\
\\
\sqrt{\displaystyle \frac{|\Delta\lambda_k|}{\lambda_k - \lambda}} \,\,\, 
\rho_{k,{\sf max}} &\quad {\lambda_{k+1}\le \lambda \le \lambda_k - |\Delta\lambda_k|}\\
\end{array}
\right.
\label{eq:density-caustic}
\ee

For $\epsilon=0.2$, the four caustic parameters can be obtained for all caustics from
the following fits (see also Fig.~\ref{fig:fitting to caustic parameters})~:
\ba
\lambda_k &=& 0.14315 \, k^{\displaystyle - 0.95986}\;\;,
\label{eq:lambda fit}
\\
\xi_k&=&  0.8169 \; + \; 0.37735 \, {\rm ln}(k)\;\;,
\label{eq:xi fit}
\\
\lambda^{"}_k &=& \left( {d^2 \lambda\over d\xi^2}\right)_k =
- 8.0272 \, k^{\displaystyle 0.9628}\;\;,
\\
\Lambda_k &=& - 0.028533 \, k^{\displaystyle - 1.9439}\;\;.
\label{eq:Lambda fit}
\ea
Similar fits can be obtained for other values of $\epsilon$.
The lower envelope (${\sf LE}$) of the density curve 
(dashed blue curve in Fig.~\ref{fig:total density}) 
is fitted by 
\be
{\rho_{\sf LE}\over \rho_b}= 0.80527 \, \lambda^{\displaystyle - 2.1093} \;\; .
\label{eq:rho_LE}
\ee
Hence the total density curve can be fitted using the above expression
together with the expressions~(\ref{eq:density-caustic}) for the density
profile $\rho_k$ of the peaks, {\it i.e.}
\be
\rho=\rho_{\sf LE}+\,\rho_k \;\; .
\label{eq:rho_total}
\ee
In the right panel of Fig.~\ref{fig:total density}, the above analytic fit is featured
by the dashed red curve while the true density~(\ref{eq:total density}) from the
numerical solution to~(\ref{eq:equation-motion}) corresponds to the solid black curve.
The fitted density profile is an excellent description of caustics even
if it differs slightly from the true density profile far from the caustics.

\begin{table}
{
\begin{tabular}{|c|c|c|c|c|}
\hline
\quad\,  $k$ \, \quad &\quad\, $\xi_k$\quad\, &\quad\, $\lambda_k$\quad\, &
\quad\, $\lambda^{\prime \prime}_k$ \quad\, 
&\quad\, $\Lambda_k$ \quad\,\\
\hline
\hline 
              1 &    0.86369 &   0.12460 &  -8.8 &  -0.0222  \\
              2 &    1.09899 &   0.06987 & -16.2 &  -0.0066 \\   
              3 &    1.24000 &   0.04899 & -23.4 &  -0.0032  \\ 
              4 &    1.34200 &   0.03782 & -30.5 &  -0.0019   \\
              5 &    1.42200 &   0.03083 & -37.6 &  -0.0012   \\
              6 &    1.48800 &   0.02604 & -44.6 &  -0.0009   \\
              7 &    1.54499 &   0.02254 & -51.6 &  -0.0007   \\
              8 &    1.59399 &   0.01988 & -58.6 &  -0.0005   \\
              9 &    1.63800 &   0.01778 & -65.5 &  -0.0004   \\
             10 &    1.67700 &   0.01608 & -72.5 &  -0.0003   \\
\hline
\end{tabular}
}
\caption{
The caustic parameters, for $\epsilon=0.2$, obtained from numerical solution
of~(\ref{eq:equation-motion}) for the first ten caustics.
Analytic fits~(\ref{eq:lambda fit})-(\ref{eq:Lambda fit}) reproduce these results
and give their values for the rest of the inner caustics.
}
\label{table:epsilon0o2}
\end{table}

%
\section{The cosmic ray signal of dark matter caustics}
\label{sec:cr_signal}
{\bf (i) Cosmic ray propagation~: the salient features}

%
%
\noindent
Supersymmetry and
extra-dimension extensions of the standard electroweak model provide
a natural candidate for dark matter in the form of a weakly interacting and massive particle
(hereafter WIMP). These species should fill up the galactic halo.
They should also annihilate in pair and produce a host of particles among
which are rare cosmic rays such as antiprotons and positrons.
The production rate
$q \left( {\mathbf x} , E \right)$ of these antimatter particles depends
on their energy $E$ and is related to the WIMP annihilation cross section
$\sigma_\mathrm{ann}$ through
\be
q \left( {\mathbf x} , E \right) = \eta \;
{\left\langle \sigma_\mathrm{ann} v \right\rangle} \,
\left\{ {\displaystyle \frac{\rho({\mathbf x})}{m_{\chi}}} \right\}^{2} \,
f(E) \;\; .
\label{source}
\ee
The coefficient $\eta$ is a quantum factor equal to $1/2$ for a
self-conjugate particle like a Majorana fermion or to $1/4$ otherwise.
In what follows, we borrow an example from supersymmetry and set
$\eta = 1/2$.
The annihilation cross section is averaged over the momenta of the
incoming DM particles to yield
${\left\langle \sigma_\mathrm{ann} v \right\rangle}$ whose value
depends on the specific microscopic interactions at stake.
The WIMP mass is denoted by $m_{\chi}$.
The energy distribution of the positrons $dN_{\ebar}/dE_{\ebar}$
or of the antiprotons $dN_{\pbar}/dE_{\pbar}$ produced in a single annihilation
is generically denoted by $f(E)$.

\vskip 0.1cm
%
%
Whatever the source mechanism, charged cosmic
rays subsequently propagate through the galactic magnetic field and are
deflected by its irregularities~: the Alfv\'en waves. In the regime where
the magnetic turbulence is strong (which is the case for the Milky Way)
cosmic ray transport needs to be investigated numerically. Monte Carlo
simulations \citep{Casse:2001be} indicate that it is similar
to space diffusion with a coefficient
\be
K(E) = K_{0} \; \beta \; {\mathcal R}^{\delta} \;\; ,
\label{K_space}
\ee
which increases as a power law with the momentum to electric charge
ratio ${\mathcal R} = {p}/{q}$ -- also called the rigidity -- of the
particle.
In addition, because the scattering centres drift inside the Milky Way with
a velocity $V_{a} \sim$ 20 to 100 km s$^{-1}$, a second order Fermi mechanism
is  responsible for some mild diffusive re-acceleration. Its coefficient $K_{EE}$
depends on the particle velocity $\beta$ and total energy $E$ and is
related to the space diffusion coefficient $K(E)$ through
\be
K_{EE} \; = \;
{\displaystyle \frac{2}{9}} \, V_{a}^{2} \,
{\displaystyle \frac{E^{2} \beta^{4}}{K(E)}} \;\; .
\ee
In the case of positrons, diffusive re-acceleration is completely dominated
by energy losses.
Finally, galactic convection wipes cosmic rays away from the disc with a velocity
$V_{C} \sim$ 5 to 15 km s$^{-1}$.

\vskip 0.1cm
%
%
After this short digest of cosmic ray transport, we can write
the master equation fulfilled by the space and energy distribution function
$\psi = dn / dE$ as
\ba
\partial_{z} \left( V_{C} \, \psi \right)  & \! - \! &  K  \Delta \psi \, + \,
\label{master_equation} \\
& \! + \! &
\partial_{E} \left\{
b^{\rm loss}(E) \, \psi  \, - \, K_{EE}(E) \, \partial_{E} \psi \right\}
= q \left( {\mathbf x} , E \right) \;\; . \nonumber
\ea
This equation applies to any species (protons, antiprotons or positrons)
as long as the rates for production $q$ and energy loss $b^{\rm loss}(E)$
are properly accounted for.
%
%
It has been solved within the framework of the semi-analytical two-zone model
which has been extensively discussed in previous works~\citep{Maurin:2001sj,Donato:2001sr} and whose
salient features we briefly recall now.
According to our approach, a steady state is assumed and the region of the
Galaxy inside which cosmic rays diffuse (the so-called diffusive halo or DH)
is pictured as a thick disc which matches the circular structure of the Milk Way
[as shown in Fig.~2 of \citep{Maurin:2002ua}]. The galactic 
disc of stars and gas,
where primary cosmic rays are accelerated, lies in the middle. It extends
radially 20 kpc from the centre and has a 
half-thickness $h$ of 100 pc. Confinement
layers where cosmic rays are trapped by diffusion lie 
above and beneath this thin
disc of gas.
The intergalactic medium starts at the vertical boundaries $z = \pm L$ as well as
beyond a radius of $r = R \equiv 20$ kpc. Notice that the half-thickness $L$
of the diffusive halo is not known and reasonable values range from 1 to 15 kpc.
The diffusion coefficient $K$ is the same everywhere whereas the convective velocity
is exclusively vertical with component $V_{C}(z) = V_{C} \; {\rm sign}(z)$. The galactic
wind, which is produced by the bulk of the disc stars like the Sun, drifts away from its
progenitors along the vertical directions, hence the particular 
form assumed here for $V_{C}$.
Notice also that the normalization coefficient $K_{0}$, the index $\delta$, the
galactic drift velocity $V_{C}$ and the Alfv\'en velocity $V_{a}$ are all unknown.
This situation may be remedied with the help of the boron to carbon B/C ratio which
is quite sensitive to cosmic ray transport and which may be used as a constraint.
The three propagation models featured in Table~\ref{tab_prop} have been borrowed
from \citet{Donato:2003xg}. The MED configuration provides the best fit to the B/C
measurements whereas the MIN and MAX models lead respectively to the minimal and
maximal allowed antiproton fluxes which can be produced by WIMP annihilation.
%
\begin{table*}
{
\begin{tabular}{|c||c|c|c|c|c|c|}
\hline
Case  & $\delta$ & $K_0$ [kpc$^2$/Myr] & $L$ [kpc] & $V_{C}$ [km/s] & $V_{a}$ [km/s] \\
\hline \hline
MIN  & 0.85 &  0.0016 & 1  & 13.5 &  22.4 \\
MED  & 0.70 &  0.0112 & 4  & 12   &  52.9 \\
MAX  & 0.46 &  0.0765 & 15 &  5   & 117.6 \\
\hline
\end{tabular}
}
\caption{
Typical combinations of diffusion parameters that are compatible with the B/C
analysis \citep{Maurin:2001sj}. As shown in \citep{Donato:2003xg}, these propagation
models correspond respectively to minimal, medium and maximal primary antiproton
fluxes.}
\label{tab_prop}
\end{table*}
%

\vskip 0.1cm
%
%
The solution of the master equation~(\ref{master_equation}) may be generically
expressed as the integral
\begin{equation}
\psi \left( \odot , E \right) =
{\displaystyle \int} \! dE_{S} \!
{\displaystyle \int}_{\rm \!\! DH} \!\!\!\!\! d^{3}{\mathbf x}_{S} \,
G \left( {\mathbf x}_{\odot} , E \leftarrow {\mathbf x}_{S} , E_{S} \right) \,
q \left( {\mathbf x}_{S} , E_{S} \right) .
\label{psi_convolution}
\end{equation}
The energy $E_{S}$ at the source runs over a range which depends on the nature
of the cosmic ray species as discussed below. The space integral is performed
over the diffusive halo. The convolution~(\ref{psi_convolution}) involves
the Green function $G$ which describes the probability for a cosmic ray
that is produced at location ${\mathbf x}_{S}$ with the energy $E_{S}$
to reach the Earth where it is detected with the degraded energy $E$.
The cosmic ray space and energy density $\psi$ can be translated into the
differential flux
$\Phi \equiv {\beta \, \psi}/{4 \pi}$ where $\beta$ stands for the particle velocity.
This flux is expressed in units of particles m$^{-2}$ s$^{-1}$ sr$^{-1}$ GeV$^{-1}$.
In the case of WIMP annihilations, it can be recast as
\be
\Phi \left( \odot , E \right) = {\mathcal F} \,
{\displaystyle \int} dE_{S} \, f(E_{S}) \,
{I}(E , E_{S}) \;\; ,
\label{phi_CR}
\ee
where the information related to particle physics has been factored out in
\be
{\mathcal F} = \eta \;
{\displaystyle \frac{\beta}{4 \pi}} \;
{\left\langle \sigma_\mathrm{ann} v \right\rangle} \,
\left\{ {\displaystyle \frac{\rho_{\odot}}{m_{\chi}}} \right\}^{2} \;\; .
\ee
The energy distribution $f(E_{S})$ describes the source spectrum and
depends on the WIMP properties. The information on the galactic DM density
profile $\rho$ as well as on the propagation of cosmic rays within the Milky Way is
summarized in the halo integral
\be
{I}(E , E_{S}) =
{\displaystyle \int}_{\rm \!\! DH} \!\!\!\!\! d^{3}{\mathbf x}_{S} \,
G \left( {\mathbf x}_{\odot} , E \leftarrow {\mathbf x}_{S} , E_{S} \right) \,
\left\{ {\displaystyle \frac{\rho({\mathbf x}_{S})}{\rho_{\odot}}} \right\}^{2}
\;\; ,
\label{I_DH_a}
\ee
where the solar neighbourhood DM density is denoted by $\rho_{\odot}$.
The halo integral ${I}(E , E_{S})$ plays a crucial role in deriving
the flux at the Earth of the antimatter species produced inside
the galactic DM halo by WIMP annihilations.
%
%
The reach of the Green function $G$ depends on the nature of the cosmic
ray particle (either positrons or antiprotons in our case) as well as on
the energies $E$ and $E_{S}$. It delineates the region of the Milky Way from
which most of the signal detected at the Earth originates. This so-called
horizon, beyond which the Green function is suppressed, plays a crucial role
in our discussion as it may or may not reach down to the centre of the Milky
Way and probe its dense DM content.

%
%
%
\begin{figure}
\centering
\includegraphics[width=\columnwidth]{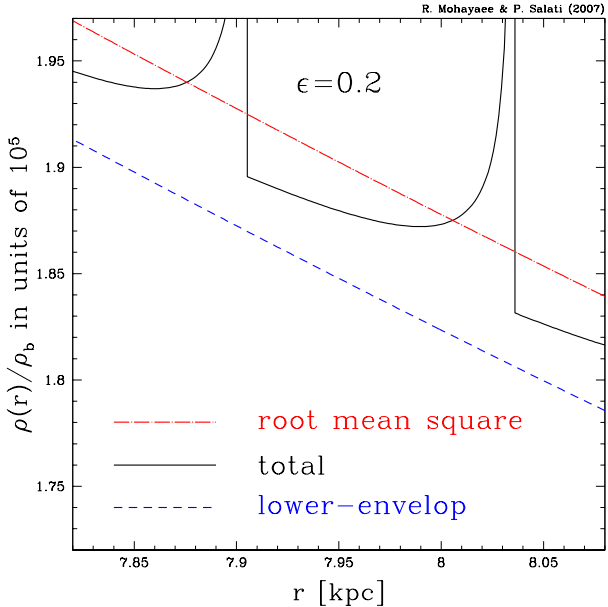}
\caption{
Typical DM density in the self-similar accretion model of
Section~\ref{sec:self_similar}. The solar system is located
at $r_{\odot} = 8$ kpc from the galactic centre and is surrounded
by two nearby thin shells.
The long-dashed dotted red curve features the root mean square density
used in the calculation of the halo integral ${I}(E , E_{S})$
whereas the short-dashed blue line corresponds to the lower envelope
$\rho_{\sf LE}$.
}
\label{fig:density_squared}
\end{figure}
%
\vskip 0.1cm
Except for high-energy positrons, the horizon is much
larger than the separation between adjacent caustics for which we find
a typical distance of $\sim$ 130 pc in the solar neighbourhood.
If the DM halo of our Galaxy results from the self-similar accretion which
we have discussed in Section~\ref{sec:self_similar}, the WIMP density that
enters in the DH integral~(\ref{I_DH_a}) is given by the sum of the lower
envelope density~(\ref{eq:rho_LE}) and the caustic
density~(\ref{eq:density-caustic}). The latter varies by several orders of
magnitude at the crossing of each shell as shown in Fig.~\ref{fig:total density}.
Because the typical scale over which the Green function $G$ changes
appreciably is much larger than the shell thickness $\Delta r_k$ and the
shell separation $r_{k} - r_{k+1}$, we may actually average the density
$\rho_{k}$ of the k$^{\rm th}$ caustic over the distance extending
from the inner radius $r_{k+1}$ to the outer boundary $r_{k}$.
This leads respectively to the linear average
\be
\overline{\rho_{k}} = \rho_{k , {\sf max}} \;
\left( 2 \sqrt{\epsilon_{k}} - \epsilon_{k} \right) \;\; ,
\label{L_average}
\ee
as well as to the quadratic average
\be
\overline{\rho_{k}^{2}} = {\rho^{2}_{k , {\sf max}}} \;
\left\{ \epsilon_{k}
\left( 1 - \ln \epsilon_{k} \right) \right\} \;\; .
\label{Q_average}
\ee
where $\epsilon_k=\Delta\lambda_k/(\lambda_k-\lambda_{k+1})$.
The DM density squared that enters in the halo integral ${I}(E , E_{S})$
may be expressed as the average
\be
\overline{\rho^{2}} \, \equiv \, \rho_{\sf LE}^{2} \; + \;
2 \, \rho_{\sf LE} \, \overline{\rho_{k}} \; + \; \overline{\rho_{k}^{2}} \;\; .
\label{DM_shell_QD}
\ee
The corresponding root mean square density is featured in
Fig.~\ref{fig:density_squared} as a function of the galactocentric
distance $r$ (long-dashed dotted red curve) together with the
actual DM density $\rho_{\sf LE} + \rho_{k}$ (solid black line)
and the short-dashed blue lower envelope $\rho_{\sf LE}$. The
turnaround radius $r_{ta}$ has been determined by requiring that the
linearly averaged DM density
$\overline{\rho} \equiv \rho_{\sf LE} + \overline{\rho_{k}}$ is equal to
a solar neighbourhood value of $\rho_{\odot} = 0.3$ GeV cm$^{-3}$.
We infer a turnaround radius $r_{ta}$ of 2.7658 Mpc.
The solar system is located between the 58$^{\rm th}$ and 59$^{\rm th}$
shells whose inner densities reach a maximum value
$\rho_{\, \sf max} = 2.13 \times 10^{7} \, \rho_{b}$
of 34.2 GeV cm$^{-3}$.
These caustics are extremely thin though, and we find
${\Delta \lambda_{58}}/(\lambda_{58} - \lambda_{59}) \sim 1.5 \times 10^{-8}$
with a shell separation of 131 pc and a shell thickness of
$\Delta \lambda_{58} = 7 \times 10^{-13}$

\vskip 0.1cm
%
%

The DM shell density derived from relation~(\ref{DM_shell_QD}) is very close
to the lower envelope density $\rho_{\sf LE}$. Both diverge at the galactic
centre with a profile index of $\gamma \simeq 2.11$.
This is physically unacceptable. The rotation velocity decreases in the bulge
as the galactocentric distance goes to zero. The self-similar model is not
a good description of the Milky Way DM halo close to its centre and a cut-off
radius $r_{\rm cut-off}$ needs to be imposed by hand in order to get a more
acceptable behaviour in that region.
The existence and the value and origin of a core at the centre of dark matter
haloes remains an open problem. The core radius claimed in the literature spans
many orders of magnitude and hence remain uncertain. It depends on very many
different parameters~: baryons, angular momentum, presence or absence of a central
black hole, DM self-annihilation, DM velocity dispersion, merger and tidal effects.
We thus present our results for a wide range of core values. The maximum range
of 3~kpc chosen here is motivated by recent studies such as those on the core radii
of dwarf galaxies (see e.g.~\citep{goerdt}).
Because baryons dominate the inner potential well of the Milky Way, the
self-similarity of the DM distribution breaks down inside the bulge. A plausible
value of $\sim$~1~kpc for the core radius $r_{\rm cut-off}$ is then set by the bulge
extension. The case of a 300 pc cut-off radius has also been investigated though
it may represent a somewhat extreme situation.

%
\begin{figure*}
\centering{
\includegraphics[width=\columnwidth]{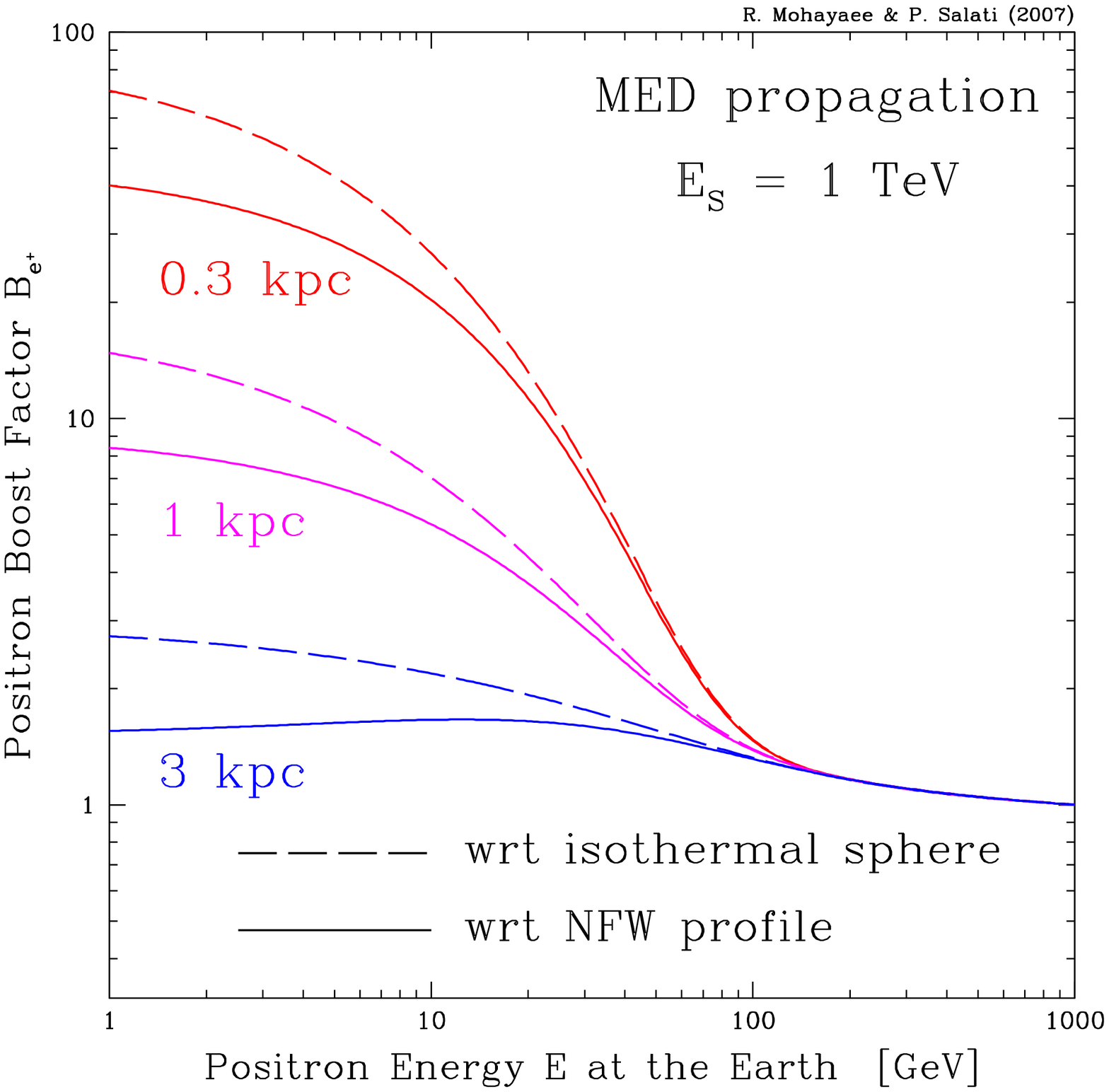}
\includegraphics[width=\columnwidth]{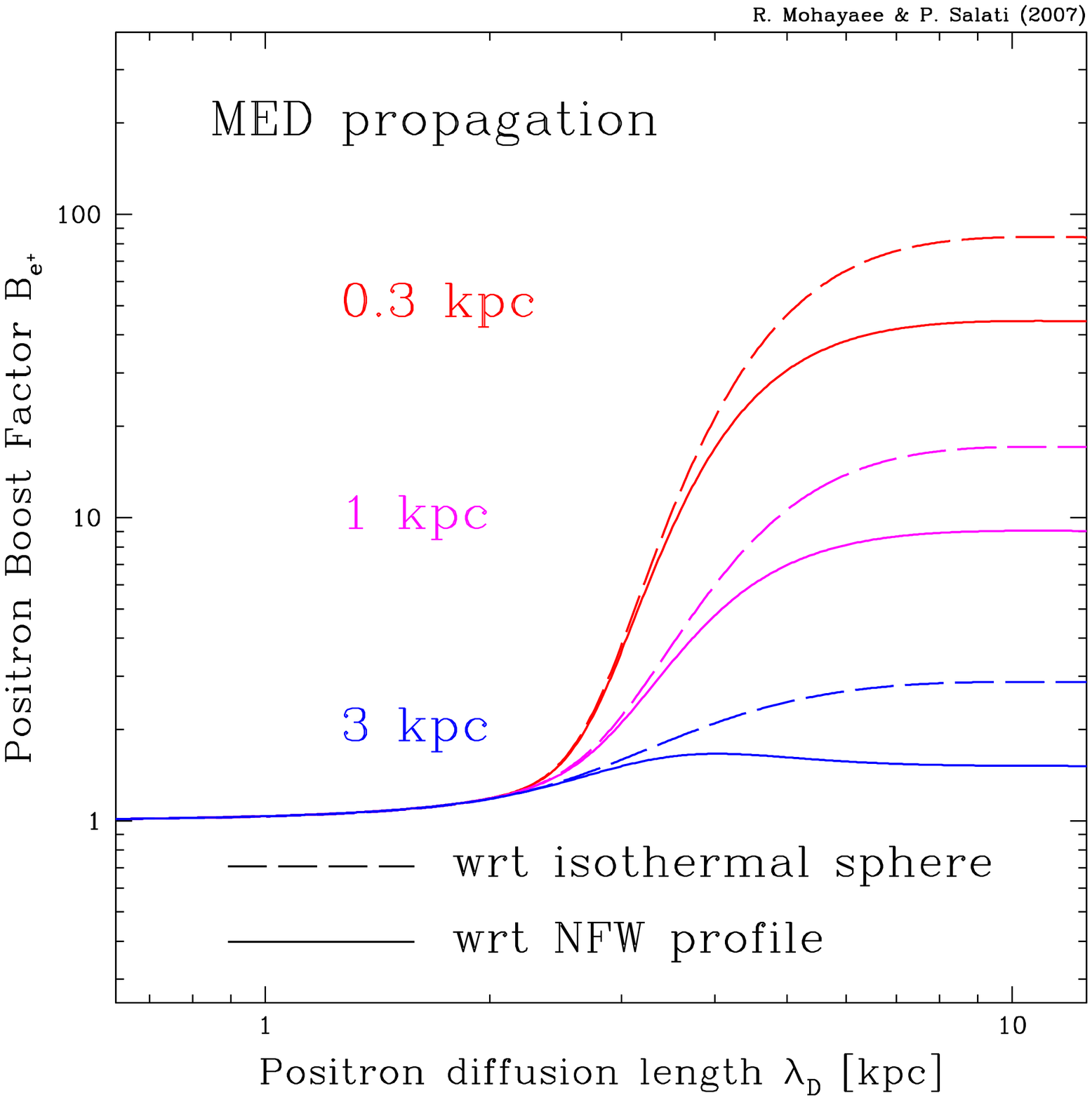}}
\caption{
Positron boost factor in the presence of DM caustics normalized to a NFW
profile (solid) or to an isothermal cored distribution (long-dashed) in
the case of the propagation model MED of Table~\ref{tab_prop}.
In the left panel, a positron line at $E_{S} = 1$ TeV has been assumed
and $\Bpos$ is plotted with respect to the energy $E$ at the Earth.
In the right panel, the boost factor is featured as a function of
the diffusion length $\lD$.
Three different values for the cut-off radius of the DM shell distribution
have been considered.
}
\label{fig:B_positron_MED}
\end{figure*}
%
\vskip 0.2cm
\noindent
{\bf (ii) The positron signal}

%
%
\noindent
The WIMPs which annihilate inside the thin and highly concentrated DM shells
of the Milky Way halo are expected to yield an a priori much higher positron
or antiproton flux at the Earth than in the more conventional case of a smooth
DM galactic distribution. Previous investigations have concentrated essentially
on the models listed in Table~\ref{tab:indices} for which the DM profile is
generically given by
\be
\rho(r) = \rho_{\odot} \,
\left\{ {\displaystyle \frac{r_{\odot}}{r}} \right\}^{\gamma} \,
\left\{
{\displaystyle \frac{1 \, + \, \left( r_{\odot} / r_{s} \right)^{\alpha}}
{1 \, + \, \left( r / r_{s}\right)^{\alpha}}}
\right\}^{\left( \beta - \gamma \right) / \alpha} \;\; .
\label{eq:profile}
\ee
The galactocentric distance $r_{\odot}$ of the solar system has been
set equal to 8 kpc.
The isothermal cored distribution~\citep{bahcall} features a constant
DM density within the inner 4 kpc whereas the Navarro, Frenk and White
profile [\citet{nfw} hereafter NFW] diverges like $r^{-1}$ at the galactic
centre. Other numerical investigations~\citep{moore} find a DM profile index
$\gamma$ of $1.16 \pm 0.14$ in good agreement with the NFW result and
should lead to the same positron or antiproton flux at the Earth.
%
\begin{table}
{\begin{tabular}{@{}|l|c|c|c|c|@{}}
\hline
Halo model & $\alpha$ & $\beta$ & $\gamma$ & $r_s$ [kpc] \\
\hline
\hline
Cored isothermal
& {\aaa} 2 {\aaa} & {\aaa} 2 {\aaa} & {\aaa} 0 {\aaa} & {\aaa} 4 {\aaa} \\
\citep{bahcall}
& {\aaa}   {\aaa} & {\aaa}   {\aaa} & {\aaa}   {\aaa} & {\aaa}   {\aaa} \\
Navarro, Frenk \& White
&        1        &        3        &        1        &        25       \\
\citep{nfw}
&                 &                 &                 &                 \\
\hline
\end{tabular}}
\caption{
Parameters for different dark matter 
density profiles [equation~(\ref{eq:profile})] for the Milky Way.
\label{tab:indices}}
\end{table}
%
In order to compare the positron signal
$\Phi_{\rm shell}\left( \odot , E \right)$ from DM caustics to the flux
$\Phi_{\rm smooth}\left( \odot , E \right)$ yielded by the DM
distributions of Table~\ref{tab:indices}, we have computed their ratio.
The positron and antiproton signals from annihilating WIMPs have already
been thoroughly investigated in the literature and shown to depend on
both the assumed DM distribution and the specific features of the
particle physics model selected to describe the DM species.
Our aim is here to gauge solely the influence of the DM shells on the
antimatter cosmic ray signal and to investigate whether or not the
annihilation of WIMPs is enhanced inside the caustics and leads
to a larger signal than in the case of an isothermal cored profile or
a NFW cusp.

%
%
\vskip 0.1cm
Because the flux depends on the source spectrum $f(E_{S})$,
we have decided to concentrate on the halo integral ${I}(E , E_{S})$
which specifically encodes information on both the DM distribution and
cosmic ray propagation. Positrons essentially diffuse on the
irregularities of the magnetic field and lose energy through synchrotron
radiation and inverse Compton scattering on the cosmic microwave background
radiation and on the galactic starlight. The energy loss rate
$b^{\rm loss}(E)$ increases with the energy $E$ as
\be
b^{\rm loss}(E) = {E_0} \, {\epsilon^2} / {\tau_E} \;\; ,
\ee
where $\epsilon \! = \! {E}/{E_0}$ and $E_{0} \! = \! 1$ GeV. The typical
energy loss timescale is $\tau_E \! = \! 10^{16}$~s and may be combined
\citep{Lavalle:2006vb,Delahaye:2007fr}
with the diffusion coefficient $K(E)$ to yield the typical diffusion length
\be
\lD = \sqrt{4 K_{0} \tilde{\tau}} \;\; .
\label{definition_lD}
\ee
The parameter $K_{0}$ is the normalization of the diffusion coefficient while
$\tilde{\tau} = \tilde{t}(E) - \tilde{t}(E_{S})$ is the typical time (including
both energy losses and diffusion) during which the positron energy decreases
from $E_{S}$ to $E$. As shown by \citet{baltz_edsjo99}, the positron energy
$E$ may be translated into the pseudo-time
\be
\tilde{t}(E) = \tau_{E} \;
\left\{
{\displaystyle \frac{\epsilon^{\delta - 1}}{1 - \delta}}
\right\} \;\; ,
\label{connection_E_pseudo_t}
\ee
where $\delta$ denotes the spectral index of the diffusion coefficient $K$
-- see relation~(\ref{K_space}).
The diffusion length $\lD$ measures actually the extension of the positron sphere.
It gauges how far positrons travel before being detected at the Earth.
A rapid inspection of equation~(\ref{connection_E_pseudo_t}) indicates that
$\lD$ increases as the detected energy $E$ decreases except for energies $E_{S}$
at the source very close to $E$. It is well known that the positron sphere is
fairly reduced at high energies, say above $\sim$ 100 GeV, whereas it
extends over several kiloparsecs below 10 GeV.

%
%
\vskip 0.1cm
The transition occurs between 10 and 100 GeV as featured in the left panel of
Fig.~\ref{fig:B_positron_MED} where the positron boost, which we define as
\be
\Bpos \equiv
{{I^{e^{+}}_{\rm shell}}(E , E_{S})}/
{{I^{e^{+}}_{\rm smooth}}(E , E_{S})} \;\; ,
\ee
is plotted as a function of the positron energy $E$ at the Earth ({\ie} after
propagation and energy loss) for a fixed value of the injection energy $E_{S}$.
Such a situation arises for Kaluza-Klein species because they can annihilate
directly into electron-positron pairs. The injection energy $E_{S}$ is equal
to the WIMP mass $m_{\chi}$ and a value of 1 TeV has been chosen here, in
agreement with universal extra dimension (UED) models. The solid and long-dashed
curves correspond respectively to the NFW profile and isothermal cored distribution
of Table~\ref{tab:indices}. Three values of the core radius $r_{\rm cut-off}$ are
displayed.
%
\begin{figure}
\centering
\includegraphics[width=\columnwidth]{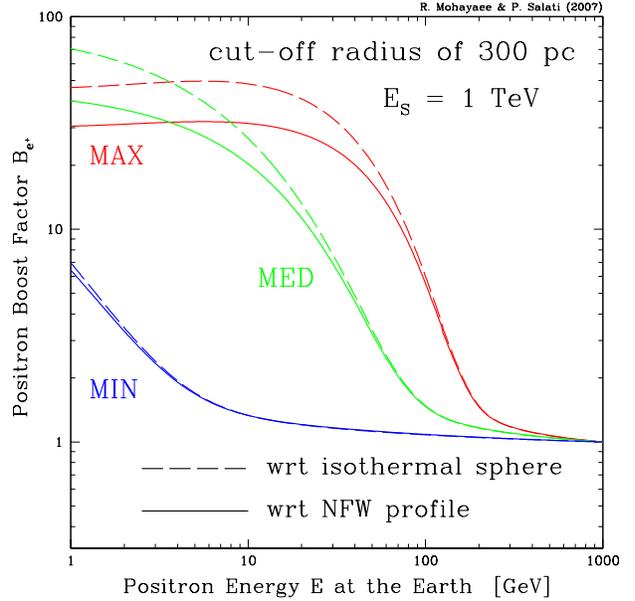}
\caption{
%
Same as in the left panel of Fig.~\ref{fig:B_positron_MED} but now for a cut-off
radius of 300 pc and for the three propagation models of Table~\ref{tab_prop}.
}
\label{fig:B_positron_300_pc}
\end{figure}
%
For energies at the Earth exceeding $\sim$ 100 GeV, the positron sphere
does not extend far away. The signal originates from the solar
neighbourhood where the DM density is equal to $\rho_{\odot}$. The halo
integrals $I^{e^{+}} \! (E , E_{S})$ are identical whatever the underlying
model for the DM density profile. As expected, the positron boost factor is
approximately equal to unity.
Below $\sim 100$ GeV, the positron sphere widens as $E$ decreases and
$\lD$ consequently gets larger. Because DM is more abundant in the shell
scenario than what is expected in the case of the smooth profiles of
Table~\ref{tab:indices}, the positron boost increases as the energy $E$
goes down from 100 GeV to 1 GeV. This trend is particularly pronounced
for the isothermal sphere (long-dashed curves) for which the DM abundance
is the smallest in the inner parts of the Milky Way. The boost factor is
shifted downwards (solid lines) by a factor of $\sim$ 2 when a NFW profile
is considered instead.
As the cut-off radius is decreased from 3 kpc down to 300 pc, the shell DM
density gets larger at the galactic centre and so does the boost.
%
Notice also that the true argument of the halo integrals $I^{e^{+}} \!$ is the
diffusion length $\lD$ which combines both energies $E$ and $E_{S}$. The right
panel of Fig.~\ref{fig:B_positron_MED}, where the positron boost $\Bpos$ is
presented as a function of $\lD$, is a mere reflection of the left plot.
Because the diffusion length increases as the energy $E$ at the Earth goes down,
these panels are indeed symmetrical.

%
%
\vskip 0.1cm
The solid and long-dashed red curves of Fig.~\ref{fig:B_positron_MED}
correspond to a cut-off radius of 300 pc. They have been plotted in
Fig.~\ref{fig:B_positron_300_pc} as a function of the positron energy
$E$ at the Earth for the three different propagation models of
Table~\ref{tab_prop}. In the MAX configuration, the spectral index
$\delta$ is the smallest and the pseudo-time $\tilde{t}(E)$ is most sensitive
to the energy $E$ as expected from relation~(\ref{connection_E_pseudo_t}).
This trend is strengthened by a large diffusion coefficient $K_{0}$.
The positron horizon probes the galactic centre as soon as the energy $E$
at the Earth gets lower than $\sim$ 50 GeV. The same regime is reached
for an energy of $\sim$ 10 GeV in the case of the MED model whereas the
positron sphere never extends to the galactic centre in the MIN case even
for energies at the Earth as low as 1 GeV.

%
%
\vskip 0.1cm

The positron line illustrates the behaviour of the horizon as a function
of the energy. This case has been discussed for pedagogical purposes though
it is inspired by Kaluza-Klein models where WIMPs annihilate directly into
electron-positron pairs. The boost can actually reach large values as $E_{S}$
is kept fixed whereas the detection energy $E$ is decreased down to a few GeV.
We should also expect a continuous positron spectrum to arise from WIMP
annihilations. Its distribution $f(E_{S})$ decreases more or less mildly with
the injection energy $E_{S}$. The signal at the Earth is then given by the
convolution of $f(E_{S})$ with the positron Green function. The positron
sphere is on average smaller than for the line and we definitely expect
smaller boost factors whose actual values depend on the steepness of the
injection spectrum.

\vskip 0.2cm
\noindent
{\bf (iii) The antiproton signature}

%
%
\noindent
The propagation of cosmic ray antiproton is dominated by diffusion.
Energy losses as well as diffusive re-acceleration do not play any major
role. A very crude approximation for the antiproton Green function
is obtained by neglecting galactic convection and solving the resulting
Poisson equation in infinite space. This yields the antiproton propagator
\be
\Gpbar \left( {\mathbf x}_{\odot} \leftarrow {\mathbf x}_{S} \right) \equiv
{\displaystyle \frac{1}{4 \pi K(E)}} \;
{\displaystyle \frac
{1}
{r_{\oplus}}} \;\; ,
\ee
where $r_{\oplus}$ denotes the distance between the Earth and the source.
The sole merit of this expression is to exhibit the importance of remote sources.
We will therefore keep in mind that the antiproton sphere is more
extended than for positrons.
The finite thickness of the diffusive halo is nevertheless a limiting factor
since cosmic rays may escape through the vertical boundaries as they wander
towards the Earth. As a consequence, the size of the antiproton sphere cannot
be much larger than the DH half-thickness $L$.
Galactic convection comes also into play. If the wind velocity $V_{C}$ is
large, cosmic rays are efficiently blown outside the Milky Way. This process
limits further the reach of the antiproton sphere.
The solution of the master equation~(\ref{master_equation}) has been
thoroughly investigated and several different techniques
\citep{Maurin:2006hy,Bringmann:2006im}
lead essentially to the same fluxes at the Earth. In the Green function
formalism, the diffusive halo is pictured as an infinite slab with no
radial boundaries. On the contrary, the Bessel expansion method takes
advantage of the axial symmetry of the propagation region and enforces
a vanishing cosmic ray flux at a distance $R = 20$ kpc from the rotation
axis of the Milky Way.
%
\begin{figure}
\centering
\includegraphics[width=\columnwidth]{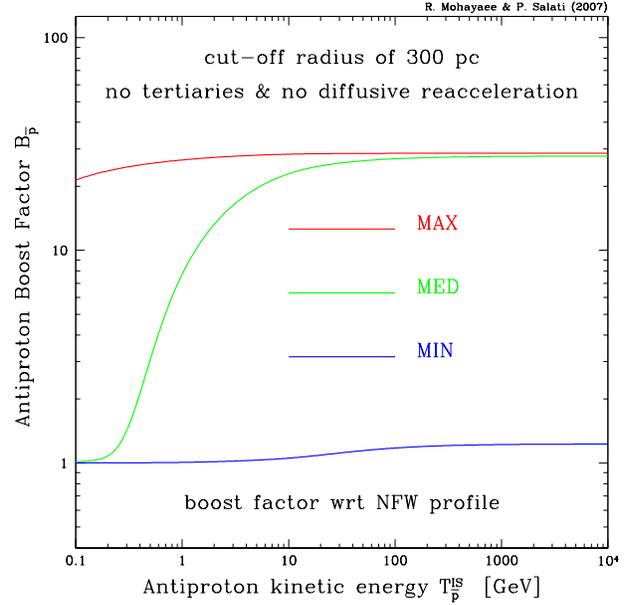}
\caption{
Antiproton boost factor $\Bpbar$ in the presence of DM caustics with respect to
a NFW profile. A cut-off radius of 300 pc has been assumed for the shell distribution.
The three propagation models of Table~\ref{tab_prop} are featured. The boost
factor is plotted as a function of the interstellar antiproton energy
$T_{\bar{\rm p}}^{\rm IS}$.
%
}
\label{fig:B_pbar_300_pc_NFW}
\end{figure}
%
%
\begin{figure*}
\centering{
\includegraphics[width=\columnwidth]{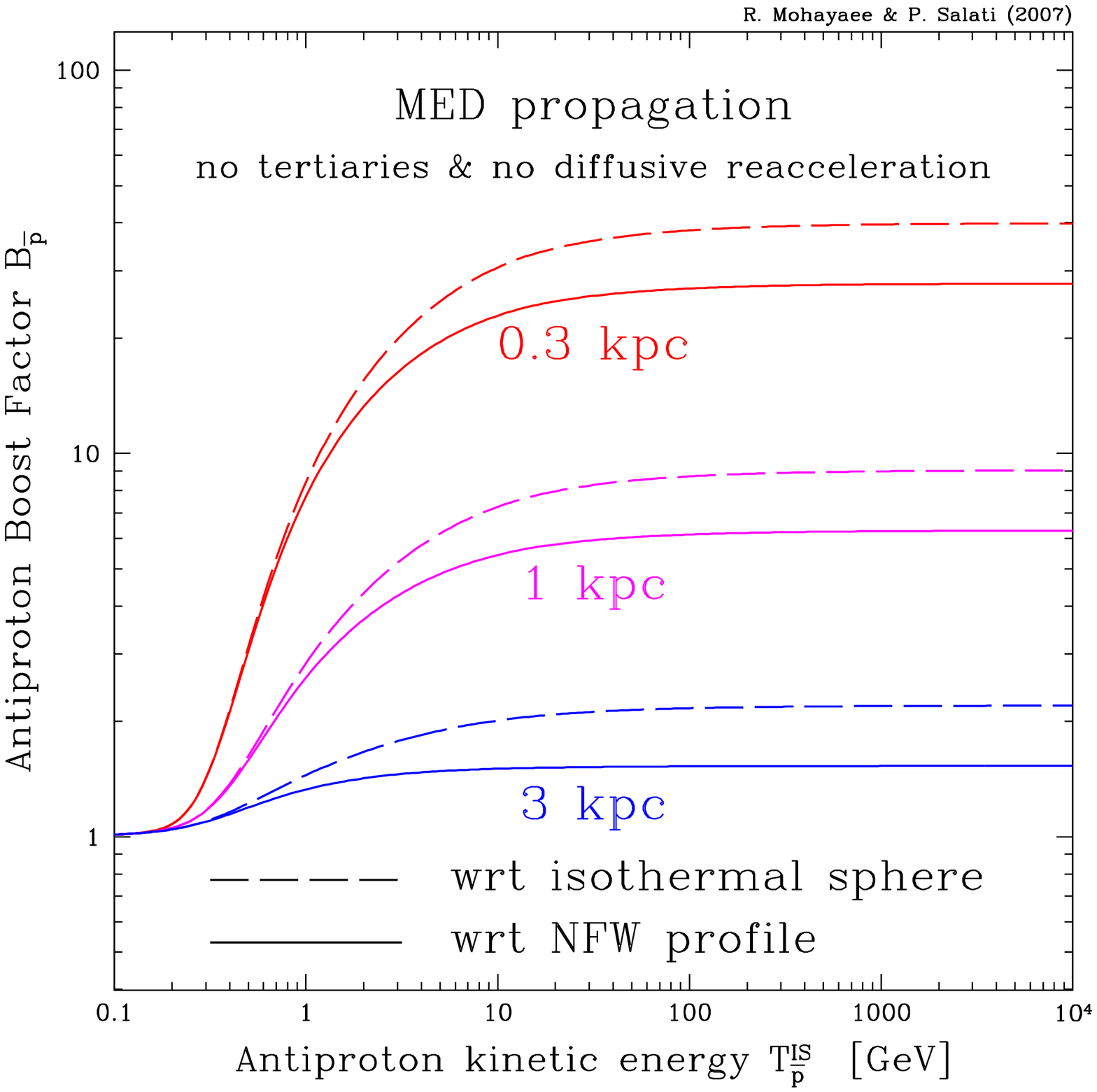}
\includegraphics[width=\columnwidth]{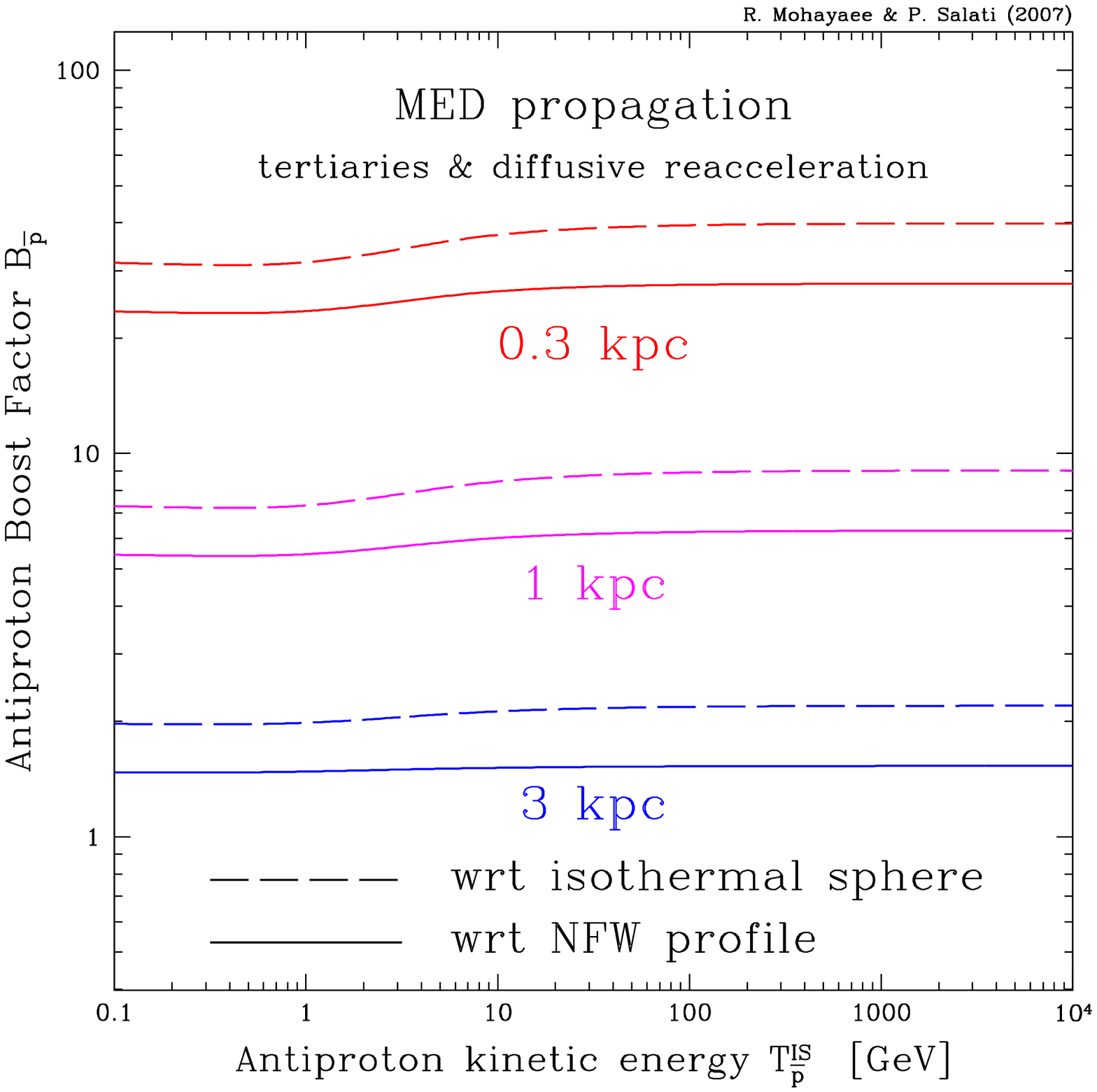}}
\caption{
Antiproton boost factor in the presence of DM caustics normalized to a NFW
profile (solid) or an isothermal cored distribution (long-dashed) in the case
of the propagation model MED of Table~\ref{tab_prop}.
Three different values for the cut-off radius of the DM shell distribution
have been considered.
Energy losses and diffusive re-acceleration have been taken into account
in the right panel as well as the production of tertiary antiprotons
which induces a dramatic increase of the boost at low energy.
}
\label{fig:B_pbar_MED}
\end{figure*}
%

%
\begin{figure}
\centering
\includegraphics[width=\columnwidth]{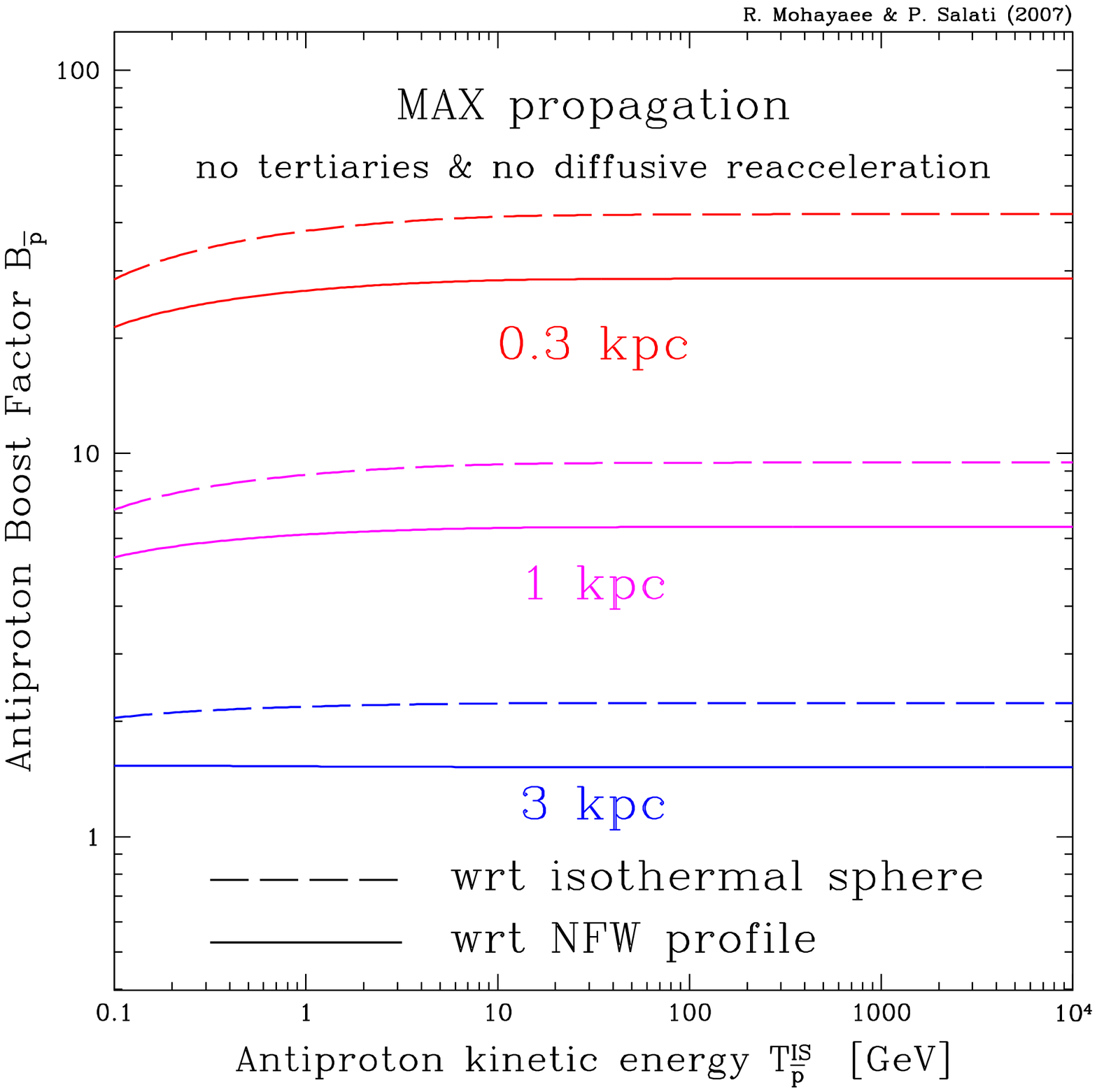}
\caption{
Same plot as the left panel of Fig.~\ref{fig:B_pbar_MED} where the
propagation model is now the MAX configuration of Table~\ref{tab_prop}.
}
\label{fig:B_pbar_MAX}
\end{figure}
%

%
%
\vskip 0.1cm
Since energy losses and diffusive re-acceleration are negligible,
the antiproton energy $E_{S}$ at the source is equal to its energy
$E$ at the Earth. Expression~(\ref{phi_CR}) may be considerably
simplified into
\be
\Phi_{\bar{\rm p}} \left( \odot , E \right) = {\mathcal F} \, f(E) \,
{I}(E) \;\; ,
\ee
where the halo integral $I$ is now a function of the antiproton energy
$E \equiv E_{S}$.
The signal generated by the DM shells of Section~\ref{sec:self_similar}
is enhanced with respect to the smooth distributions of Table~\ref{tab:indices}
by the boost factor
\be
\Bpbar \equiv 
{{I^{\bar{\rm p}}_{\rm shell}}(E)}/
{{I^{\bar{\rm p}}_{\rm smooth}}(E)} \;\; .
\ee
The latter is plotted in Fig.~\ref{fig:B_pbar_300_pc_NFW}
as a function of the antiproton interstellar kinetic energy
$T_{\bar{\rm p}}^{\rm IS}$ for the three propagation models
of Table~\ref{tab_prop}. A cut-off radius of 300 pc has been assumed
for the DM shell distribution. Its boost $\Bpbar$ is calculated with
respect to a smooth NFW halo.
The Green function technique and the Bessel expansion method yield
identical results.
Energy losses and diffusive re-acceleration have been switched off
as well as the production of tertiary antiprotons.
The MAX propagation model (red curve) is characterized by a large
value of the half-thickness $L$ and a small convection velocity
$V_{C}$. The antiproton sphere spreads therefore over a large portion
of the Milky Way and reaches the inner dense regions of its DM halo.
The boost factor is essentially constant and is equal to $\sim 30$
above a few GeV.
For exactly the opposite reasons, the MIN configuration (blue line)
does not lead to any enhancement of the antiproton signal in the
presence of DM caustics. The diffusive halo is too thin and the galactic
wind too strong to let the antiprotons originating from the galactic
centre to reach the Earth.
The MED case (in green) features the intermediate situation. Above
$\sim$ 100 GeV, diffusion takes over galactic convection and the
antiproton sphere is essentially limited by the half-thickness $L$
of the DH. That sphere extends sufficiently close to the centre to
become sensitive to its large DM density, hence a boost factor identical
to the MAX result. Below $\sim$ 10 GeV, the galactic wind is on the
contrary strong enough to restrain the antiproton sphere from spreading
and the MIN situation is recovered.
%
\begin{figure*}
\centering{
\includegraphics[width=\columnwidth]{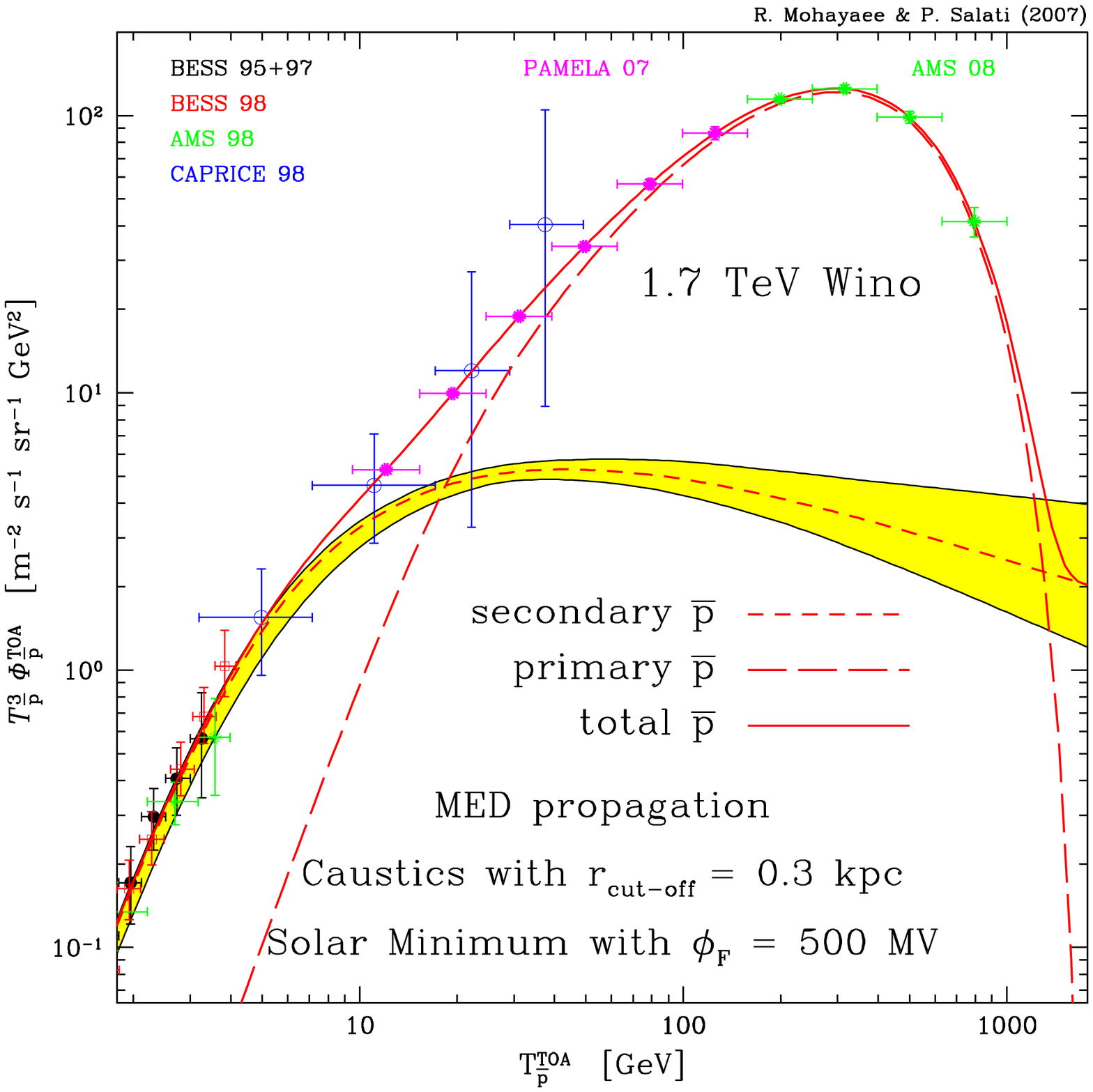}
\includegraphics[width=\columnwidth]{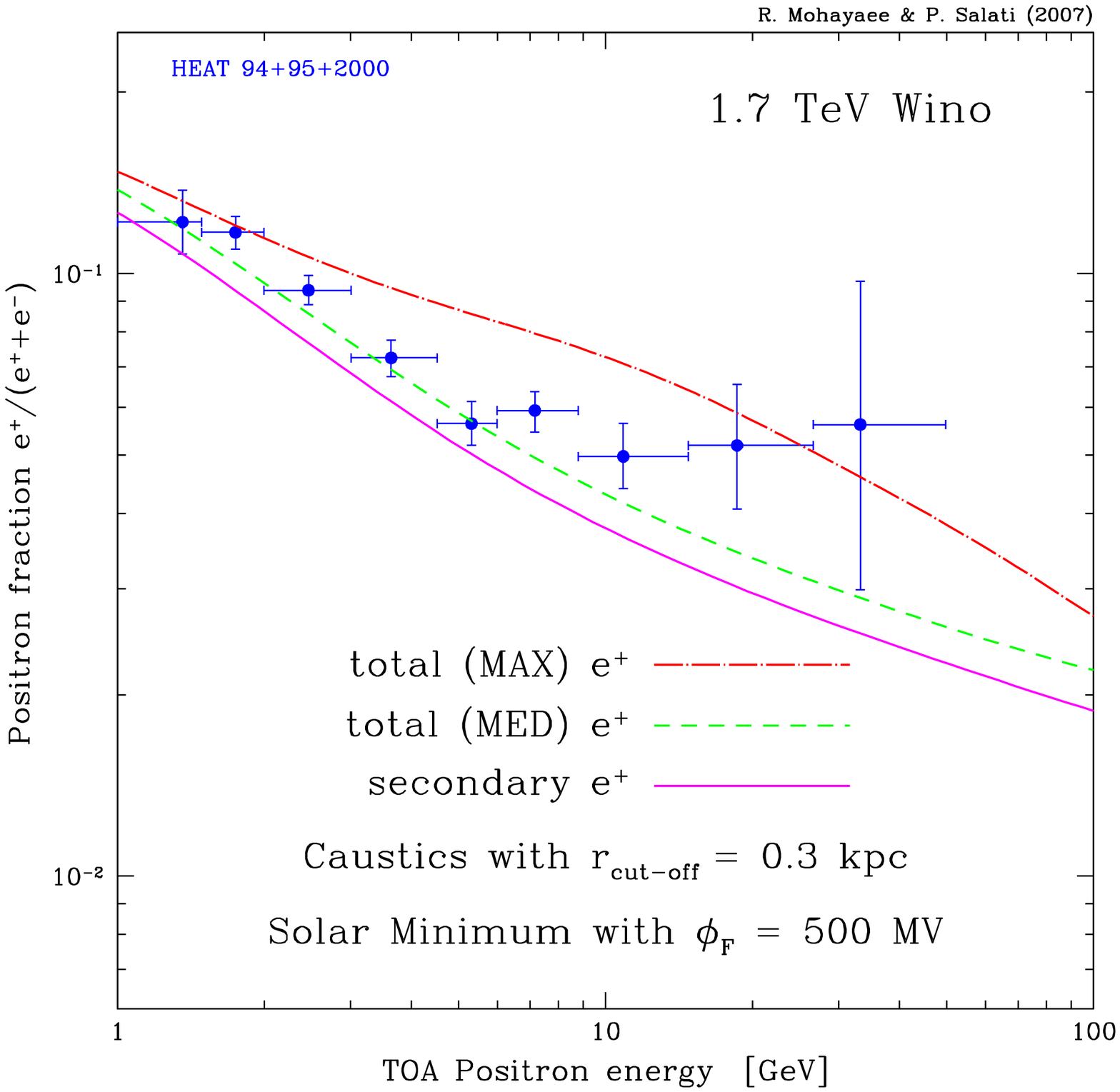}}
\caption{
Signals expected in the case of a 1.7 TeV Wino and a DM shell distribution with
a cut-off radius of 300 pc.
In the left panel, the antiproton flux
$T_{\bar{\rm p}}^{3} \, \Phi_{\bar{\rm p}}^{\rm TOA}$
is plotted as a function of the antiproton kinetic energy $T_{\bar{\rm p}}^{\rm TOA}$
and compared to several measurements
\citep{Orito:1999re,Maeno:2000qx,Boezio:2001ac,Aguilar:2002ad}
and projected observations
\citep{Picozza:2006nm,Barao:2004ik}.
The MED propagation model has been assumed.
In the right panel, the positron fraction is featured for the MAX and MED
propagation sets together with the HEAT data
\citep{heat}.
}
\label{fig:flux_WINO}
\end{figure*}
%

%
%
\vskip 0.1cm
In Fig.~\ref{fig:B_pbar_MED}, the MED propagation mode has been selected.
The boost $\Bpbar$ is featured as a function of the antiproton interstellar
kinetic energy $T_{\bar{\rm p}}^{\rm IS}$ for three different values
of the cut-off radius $r_{\rm cut-off}$. It has been normalized to a NFW
profile (solid) or an isothermal cored distribution (long-dashed). The
solid red curve of the left panel corresponds therefore to the green line
of Fig.~\ref{fig:B_pbar_300_pc_NFW} and the decrease of the boost at low
energy can be explained as a result of the galactic wind taking over
diffusion.
In the right panel, the Bessel expansion technique alone has been used in
order to implement energy losses and diffusive re-acceleration. The inelastic
and yet non-annihilating interaction of an antiproton with an hydrogen atom
of the interstellar medium is possible through the production of a $\Delta$
resonance which subsequently decays into a proton and a pion. The antiproton
is not destroyed but loses a significant portion of its energy in a single
collision. This so-called tertiary process has also been taken into account.
The curves are significantly flattened since the low-energy
tail of the spectrum is replenished by higher energy antiprotons undergoing
these inelastic and non-annihilating scatterings.
Since the energy $E$ drifts now with respect to its initial value $E_{S}$,
the injected spectrum matters. In the right panel, $f(E_{S})$ has been
assumed to be constant. This is an unrealistic situation as the antiproton
spectrum generated by WIMP annihilation is expected to be softer than
a flat distribution. The panels of Fig.~\ref{fig:B_pbar_MED} feature therefore
two extreme trends and the actual boost should follow an intermediate
behaviour.

%
%
\vskip 0.1cm
In Fig.~\ref{fig:B_pbar_MAX}, energy losses and diffusive re-acceleration
have been switched off as well as the above mentioned tertiary process.
The situation is therefore similar to what has been assumed in the left
panel of Fig.~\ref{fig:B_pbar_MED} up to the propagation model which is
now the MAX configuration of Table~\ref{tab_prop}. The right panel
of Fig.~\ref{fig:B_pbar_MED} (MED propagation with tertiaries and diffusive
re-acceleration) is similar to Fig.~\ref{fig:B_pbar_MAX} (MAX propagation
without tertiaries nor diffusive re-acceleration). We conclude that in the
MED configuration, the antiproton horizon may be somewhat stretched at low
energy by the tertiary process which allows a fraction of the high energy
antiprotons to make it from the inner Galaxy to the Earth where they
subsequently populate the energy band below $\sim$ 10 GeV.

%
%
\vskip 0.1cm
The annihilation cross section of WIMPs is in general set by the requirement
that their relic abundance today should be compatible with the WMAP
measurement of $\Omega_{M} \sim 0.25$. This leads to a value of
${\left\langle \sigma_\mathrm{ann} v \right\rangle}$
of order a few $10^{-26}$ cm$^{3}$ s$^{-1}$ and to an undetectable signal.
The presence of substructures in the Milky Way halo has been invoked to
increase the cosmic ray fluxes at the Earth by an amount which is still
debated \citep{Lavalle_Maurin}.
The aim of this article is actually to estimate that value in the
case of caustics. We find that the boost hardly exceeds $\sim$ 30. With such
a value and assuming the above mentioned annihilation cross section, the
antiproton and positron signals are barely detectable.
There is however a case where
${\left\langle \sigma_\mathrm{ann} v \right\rangle}$ could be as large as
$10^{-24}$ cm$^{3}$ s$^{-1}$. In an extensive analysis of a fairly generic
supersymmetric model, \citet{Profumo:2004at} have shown that a heavy Wino-like
LSP often appears as a plausible DM candidate. This species is expected for
example in anomaly mediated supersymmetry breaking (AMSB) scenarios
\citep{Ullio:2001qk} and relic density requirements point towards a mass of
1.7 TeV. We emphasize that this configuration is quite generic. 
Non-perturbative, binding energy effects imply a greatly enhanced annihilation
cross section today, when the Winos have very small galactic velocities
as shown by \citet{Hisano:2004ds}. In Fig.~\ref{fig:flux_WINO}, a value of
$1.02 \times 10^{-24}$ cm$^{3}$ s$^{-1}$ has been assumed for
${\left\langle \sigma_\mathrm{ann} v \right\rangle}$. In this limit, heavy
Winos annihilate almost exclusively into gauge bosons, with a branching ratio
of 80\% into $W^{+} W^{-}$ pairs and of 20\% into $Z^{0} Z^{0}$ pairs in
our example.
In the left panel, the antiproton spectrum is plotted as a function of the
antiproton kinetic energy. The MED propagation mode is selected here and 
since in that case the antiproton horizon is fairly limited at low energy,
the yield from DM annihilation is small and compatible with the measurements.
Above $\sim$ 10 GeV, the signal however increases dramatically with respect
to the background (yellow band) which consists of secondary antiprotons. The
latter are produced conventionally by the spallations of high-energy CR nuclei
on the interstellar gas. The solid red curve is even compatible with the CAPRICE
data \citep{Boezio:2001ac} which possibly hint at a spectral distortion.
The largest discrepancy between the DM shell signal and the secondary
background is reached for an energy of 300 GeV. The antiproton sphere
clearly reaches the core of the DM shell distribution.
Notice that no artificial enhancement of the antiproton primary flux
is necessary here. We have used solely the DM distribution resulting from
the self-similar infall model of Section~\ref{sec:self_similar} with
no extra boost factor. Our only assumption is a cut-off radius of 0.3 kpc.
Should we have considered larger core radii, we would have obtained however
a smaller signal.
In the right panel, the positron fraction is presented as a function of
the positron energy $E$ at the Earth. The solid magenta curve corresponds
to the background alone as derived by \citet{strongmoska}. Secondary positrons
are produced in the Galaxy from the collisions of cosmic ray proton
and helium nuclei on the interstellar medium.
The short-dashed green line features the signal of our 1.7 TeV Wino
and is based on the MED propagation model. The positron sphere is sensitive to the
high DM density of the central caustics only at low energy, in a regime
where the secondary background is large enough to hide easily a spectral distortion.
This is no longer the case if the MAX propagation model is substituted for the
MED configuration. The positron sphere becomes large enough to reach the
galactic centre below 50 GeV and the long-dashed red curve exhibits a clear
excess with respect to the background. The HEAT measurements \citep{heat} are
not well reproduced though. They point towards a positron excess which increases
above 10 GeV, a trend which our shell density~(\ref{DM_shell_QD}) clearly
does not match.
A Kaluza-Klein DM species with a substantial positron line has been invoked
to fit the HEAT distortion \citep{Cheng:2002ej,Hooper:2004bq}. However, an
artificial boost factor of $5 \times (0.43/0.3)^{2} \sim 10$ on the entire
positron energy range is necessary.
Another possibility relies on the presence of high concentrations of WIMPs
in the solar neighbourhood. The increase of the HEAT excess 
with energy indicates
that a local mechanism is at play. The probability for a DM clump to be close
to the Earth has been investigated \citep{Hooper:2003ad,Lavalle:2006vb}.
In our case, the caustics themselves could play the same role as the clumps,
because the DM density reaches there a priori extreme values. We may wonder
then how large the positron signal could be, should the solar system
be close to one of these shells.

%
\section{Dark matter draperies and the positron excess}
\label{sec:draperies}

%
%
Unlike antiprotons which have a very long diffusion length, the positrons
that are detected at the Earth before significantly losing their
energy ($E \simeq E_{S}$) have covered on average a small distance. For
this reason, they can be important tracers of nearby density peak structures
such as the dark matter ``draperies" on which we focus here.
As mentioned in Section~\ref{sec:cr_signal}, the solar system is located
between two nearby shells. In our self-similar infall model, the 58$^{\rm th}$
and 59$^{\rm th}$ caustics lie respectively at a distance of 8.0348 and 7.9040 kpc
from the galactic centre. The shell separation $e = r_{58} - r_{59}$ is equal
to 130.8 pc. It is 8 orders of magnitude larger than the shell thickness for
which a typical value is given by $\Delta r_{58} = 5.9 \times 10^{7}$ km,
hence the term ``drapery". A turnaround radius $r_{ta}$ of 2.7658 Mpc has
been obtained by setting the linearly-averaged density
$\overline{\rho} = \rho_{\sf LE} + \overline{\rho_{58}}$ equal to
the solar neighbourhood value $\rho_{\odot} = 0.3$ GeV cm$^{-3}$.
The density $\rho_{58}$ of the 58$^{\rm th}$ caustic reaches a maximum
of $\rho_{58,{\sf max}}~=$~34.2 GeV cm$^{-3}$ inside an homogeneous slab
whose thickness $\Delta r_{58}$ is equal to the distance between the Sun
and Mercury. Should the Earth wander in this region of highly concentrated DM,
the positron signal would be enhanced by a factor of
$(\rho_{58,{\sf max}} / \rho_{\odot})^{2} \sim 13,000$, hence the importance
of exploring this possibility in greater detail.

%
%
\vskip 0.1cm
To start, we note that changing slightly the value of $\rho_{\odot}$
modifies the turnaround radius and affects the relative position of the solar
system and the DM shells. The Earth may lie within the densest region of a
caustic but the probability for such a situation is grossly given
by the ratio $\epsilon_{58} = \Delta r_{58} / e \sim 1.5 \times 10^{-8}$ and
is vanishingly small.

%
%
\vskip 0.1cm
Playing the devil's advocate, let us nevertheless assume that the solar
system is indeed embedded inside this density peak region with typical DM
density $\rho_{58,{\sf max}}$.
Should this be the case, the positron signal would still be so smeared
by diffusion and energy losses that it would be marginally enhanced with
respect to the results presented in Section~\ref{sec:cr_signal}. Even in
the extreme situation of a line where the injection energy is
$E_{S} = m_{\chi}$, positrons are detected at energies $E < E_{S}$ and
originate from a sphere all the more extended as the energy difference
$\Delta E = E_{S} - E$ is large. Let us consider the MED propagation model.
The spectral index of the diffusion coefficient $K$ is $\delta = 0.7$.
In the limit where the energy difference $\Delta E$ is small compared
to the positron energy $E \simeq E_{S}$, we may differentiate
relation~(\ref{connection_E_pseudo_t}) and combine the result with
definition~(\ref{definition_lD}) to get
\be
\lD = 3.77 \; {\rm kpc} \;
{\Delta \epsilon}^{1/2} \; \epsilon_{S}^{- 0.65} \;\; ,
\label{lD_close_line}
\ee
where $\epsilon_{S} \! = \! {E_{S}}/{E_0}$ and $E_{0} \! = \! 1$ GeV.
For a typical positron energy of 100 GeV, the diffusion length is equal
to
\be
\lD = 1.89 \; {\rm kpc} \;
\sqrt{\displaystyle \frac{\Delta E}{E_{S}}} \;\; ,
\label{lD_line_a}
\ee
Because the experimental determination of the energy $E$ is performed with
a limited accuracy, positrons emitted monochromatically at $E_{S} = 100$
GeV are collected in an energy band set by the resolution of the
instrument. Even in the optimistic case where energy is measured at the 1\%
level, {\ie} with ${\Delta E}/{E_{S}} = 10^{-2}$, relation~(\ref{lD_line_a})
yields a positron sphere which reaches as far as 190 pc from the Earth
at $E$ = 99 GeV. The signal collected in the energy bin of the line does
not originate solely from the densest part of the caustic. It has been
produced from a much more extended zone whose average DM density is
\be
\overline{\rho} \simeq \rho_{\sf LE} \simeq \rho_{\odot}
= 0.3 \; {\rm GeV \; cm^{-3}} \;\; .
\ee
Remember that if the shells are dense, they are also extremely thin.
The DM density $\rho_{58}$ for instance undergoes a discontinuity at
the external boundary of the 58$^{\rm th}$ caustic.
It also drops violently inwards, decreasing from 34.2 down to
0.3~GeV~cm$^{-3}$ on a distance of $2.5 \times 10^{-2}$~pc only.
Averaged over a distance $D$, the 58$^{\rm th}$ caustic density
is given by
\be
\overline{\rho_{58}} = 2 \; \rho_{58,{\sf max}} \;
\sqrt{\displaystyle \frac{\Delta r_{58}}{D}} \;\; ,
\ee
where relation~(\ref{L_average}) has been used. For a distance $D = 1$ pc,
we get a value of 0.095 GeV cm$^{-3}$ which barely represents a third of
the solar neighbourhood value. Averaged now over the distance $e$ separating
two nearby shells, it amounts to 2.8\% of the density $\rho_{\odot}$,
in agreement with the relative positions of the long-dashed dotted red
curve (averaged total DM density) and the short-dashed blue line
(lower envelope) of Fig.~\ref{fig:density_squared}.

%
%
In order to elaborate on the smearing of the positron horizon by the
limited resolution of energy measurements, some modelling is necessary.
Since the shells are extremely thin, we will replace the actual DM density
by the fine-grained distribution $\rho_{\sf eff}$ where
\be
{\displaystyle \frac{\rho_{\sf eff}^{2}}{\rho_{\odot}^{2}}} =
{\displaystyle \frac{\rho_{\sf LE}^{2}}{\rho_{\odot}^{2}}}
\; + \;
{\displaystyle \sum_{{\rm shells} \; k}} \,
A \, e \, \delta \left( z - z_{k} \right) \;\; .
\label{fine_grained}
\ee
%
The shells are now pictured as infinitely thin slabs located at the positions
$z = z_{k}$ on the radial axis connecting the galactic centre to the Earth.
They are separated from each other by the distance $e$ defined above.
Setting the total number of DM annihilations constant, the parameter $A$ is
found to be equal to
\be
A =
{\displaystyle \frac{\overline{\rho^{2}} - \rho_{\sf LE}^{2}}{\rho_{\odot}^{2}}}
\simeq 2 \; {\displaystyle \frac{\overline{\rho_{58}}}{\rho_{\odot}}} \;\; ,
\ee
with a typical value of $0.0554$.
The effect which we discuss here is the strongest in the case of
a positron line for which the injection spectrum is
$f(E_{S}) = \delta \left( E_{S} - m_{\chi} \right)$. The positron
flux at the Earth is given by the product
\be
\Phi_{e^{+}} \! \left( \odot , E \right) = 
{\mathcal F} \; {I^{e^{+}}} \! (E , E_{S} = m_{\chi}) \;\; ,
\ee
where the halo integral ${I^{e^{+}}}$ is the convolution~(\ref{I_DH_a})
between the DM density squared
and the positron Green function $G_{e^{+}}$. Close to the line energy, for
$E \simeq m_{\chi}$, the positron horizon is so limited that we may not replace
the actual DM density by its coarse-grained average~(\ref{DM_shell_QD}), especially
if the Earth lies in the region where the caustic density is the highest.
The results presented in Section~\ref{sec:cr_signal} should potentially
vary if we compute now the halo integral ${I^{e^{+}}}$ with the
fine-grained density~(\ref{fine_grained}).
%
The relative change in the positron flux is equal to the ratio
\be
{\displaystyle \frac{\delta {I^{e^{+}}}}{I^{e^{+}}}} \equiv
{\displaystyle \frac{I^{e^{+}}_{\rm fine} - I^{e^{+}}_{\rm coarse}}{I^{e^{+}}_{\rm coarse}}}
\;\; .
\ee
Because the positron propagator may be expressed as
\be
G_{e^{+}} \left( \mathbf{x} , E \leftarrow \mathbf{x}_{S} , E_{S} \right) =
{\displaystyle \frac{\tau_{E}}{E_{0} \, \epsilon^{2}}} \;
\tilde{G} \left( \mathbf{x} \leftarrow \mathbf{x}_{S} ; {\lD} \right)
\;\; ,
\label{positron_propagator}
\ee
the halo integral ${I^{e^{+}}}$ is related to the convolution $\tilde{I}$
of the DM density squared with the heat Green function $\tilde{G}$ through
\be
{I^{e^{+}}} \! (E , E_{S}) \equiv
{\displaystyle \frac{\tau_{E}}{E_{0} \, \epsilon^{2}}} \; \tilde{I}(\lD) \;\; ,
\ee
where
\be
\tilde{I}(\lD) =
{\displaystyle \int}_{\rm \!\! DH} \!\!\!\!\! d^{3}{\mathbf x}_{S} \;
\tilde{G} \left( {\mathbf x}_{\odot} \leftarrow \mathbf{x}_{S} ; {\lD} \right) \;
\left\{ {\displaystyle \frac{\rho({\mathbf x}_{S})}{\rho_{\odot}}} \right\}^{2}
\;\; .
\ee
Considering the fine-grained DM distribution~(\ref{fine_grained}) instead of
the coarse-grained average~(\ref{DM_shell_QD})
induces a relative change in the positron flux equal to
$\delta \tilde{I} / \tilde{I}_{\rm coarse}$ where
\be
\delta \tilde{I} \equiv \tilde{I}_{\rm fine} - \tilde{I}_{\rm coarse}
\;\; ,
\ee
and
\be
\tilde{I}_{\rm coarse} =
{\displaystyle \int}_{\rm \!\! DH} \!\!\!\!\! d^{3}{\mathbf x}_{S} \;
\tilde{G} \left( {\mathbf x}_{\odot} \leftarrow \mathbf{x}_{S} ; {\lD} \right) \;
{\displaystyle \frac{\overline{\rho^{2}}}{\rho_{\odot}^{2}}}
\;\; .
\ee
We are interested here in the limit where the positron diffusion length $\lD$ is
small compared to the intershell separation $e~\sim~131$~pc. In this regime,
$\lD$ is much smaller than the DH half-thickness $L$ and the heat Green function
$\tilde{G}$ simplifies into \citep{baltz_edsjo99,Lavalle:2006vb}
\begin{equation}
\tilde{G} \left( {\mathbf x} \leftarrow \mathbf{x}_{S} ; {\lD} \right) =
{\displaystyle \frac{1}{\pi^{3/2} \lD^{3}}} \;
\exp \left\{ - \,
{\displaystyle \frac{(\mathbf{x} - \mathbf{x}_{S})^{2}}{\lD^{2}}} \right\}
\;\; .
\label{propagator_reduced_3D}
\end{equation}
Since $\overline{\rho^{2}} \simeq \rho_{\odot}^{2}$, we are led to the conclusion
that $\tilde{I}_{\rm coarse} \simeq 1$.
Our task consists then in evaluating the difference
\be
\delta \tilde{I}({\lD}) =
{\displaystyle \int}_{\rm \!\! DH} \!\!\!\!\! d^{3}{\mathbf x}_{S} \;
\tilde{G} \left( {\mathbf x}_{\odot} \leftarrow \mathbf{x}_{S} ; {\lD} \right) \;
\left\{ {\displaystyle
\frac{\rho_{\sf eff}^{2} - \overline{\rho^{2}}}{\rho_{\odot}^{2}}}
\right\}
\;\; ,
\ee
and in comparing it to unity. With our fine-grained expression~(\ref{fine_grained}),
this integral becomes
\be
\delta \tilde{I}({\lD}) =
{\displaystyle \int}_{\rm \!\! DH} \!\!\!\!\! d^{3}{\mathbf x}_{S} \;
\tilde{G} \; A \,
\left\{ - 1 \, + \,
{\displaystyle \sum_{{\rm shells} \; k}} \, e \, \delta \left( z - z_{k} \right)
\right\}
\;\; .
\ee
To proceed further, we simplify the Gaussian expression~(\ref{propagator_reduced_3D})
for the heat propagator $\tilde{G}$ into the step function
\be
\tilde{G} =
{\displaystyle \frac{1}{V_{D}}} \;
\theta \left( \lD - r_{\oplus} \right) \;\; ,
\ee
where $r_{\oplus}$ denotes the distance of the source to the Earth.
The radius of the positron sphere is $\lD$ and its volume is
$V_{D} = {4 \pi {\lD}^{3}}/{3}$. We readily find that
\be
\delta \tilde{I}({\lD}) = A \;
\left\{ - 1 \, + \!\!\!
{\displaystyle
\sum_{{- \lambda_{\rm D}} \leq z_{k} \leq {\lambda_{\rm D}}}}
{\displaystyle \frac{\pi \, e}{V_{D}}} \,
\left( {\lD}^{2} - z_{k}^{2} \right)
\right\} \;\; ,
\ee
where the sum runs over the DM shells that intersect the positron sphere.
In the limit where the positron diffusion length $\lD$ is much larger
than the caustic separation $e$, the difference $\delta \tilde{I}$
vanishes as expected. In this regime, the positron sphere is so large
that the coarse-grained density~(\ref{DM_shell_QD}) is enough to describe
appropriately the DM distribution.
We are actually interested here in the opposite limit where the ratio
${\lD}/e$ is small and where the draperies come into play. The difference
$\delta \tilde{I}$ may be conveniently expressed as
\be
\delta \tilde{I}({\lD}) = A \;
{\displaystyle \frac{e}{\lD}} \; {\mathcal D} \;\; ,
\ee
where the drapery function ${\mathcal D}$ is defined as
\be
{\mathcal D} \equiv
- {\displaystyle \frac{\lD}{e}} \, + \!\!\!
{\displaystyle
\sum_{{- \lambda_{\rm D}} \leq z_{k} \leq {\lambda_{\rm D}}}}
{\displaystyle \frac{3}{4}} \,
\left(1 -
{\displaystyle \frac{z_{k}^{2}}{{\lD}^{2}}}
\right) \;\; .
\ee
%
\begin{figure}
\centering
\includegraphics[width=\columnwidth]{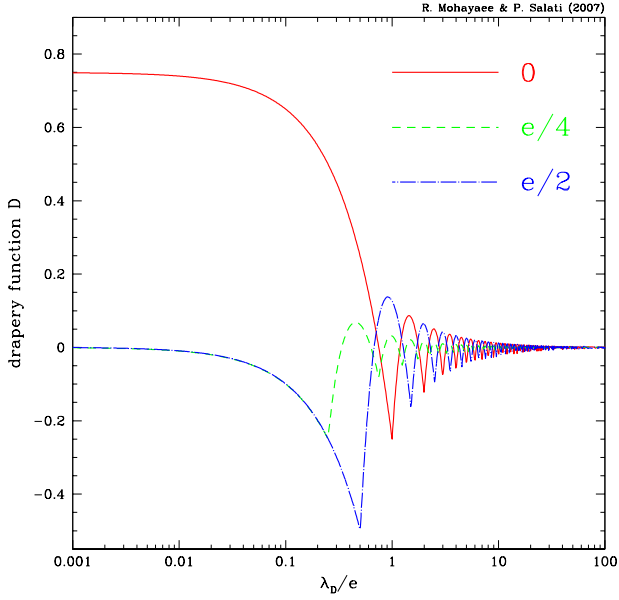}
\caption{
The drapery function ${\mathcal D}$ is plotted against the ratio ${\lD}/{e}$
for three different values of the distance to the nearest DM shell.
The solid red curve features the case where the Earth is located exactly inside
a drapery and where the DM density is the highest. The integral $\delta \tilde{I}$
is maximal.
The long-dashed dotted blue line corresponds to the opposite situation where
the Earth lies at mid-distance between two caustics. The integral
$\delta \tilde{I}$ is minimal with a negative value.
The intermediate situation for which the nearest shell is located at a distance
of $e/4$ leads to the short-dashed green curve.
}
\label{fig:drapery_function}
\end{figure}
%
The latter is plotted in Fig.~\ref{fig:drapery_function} as a function of the ratio
${\lD}/{e}$.
The minimal value of the difference $\delta \tilde{I}$ is reached when no DM shell
intersects the positron sphere. The Earth lies between two nearby draperies. This
case is featured for instance by the short-dashed green and long-dashed dotted blue
curves of Fig.~\ref{fig:drapery_function} and provided that the positron diffusion
length $\lD$ is small enough. The minimal value of $\delta \tilde{I}$ is equal to $- A$
and is negligible with respect to $\tilde{I}_{\rm coarse} \simeq 1$.
More complicated is the maximal configuration for which the Earth lies exactly inside
a drapery and where the DM density is the highest. This case corresponds to the
solid red line of Fig.~\ref{fig:drapery_function}. For very small values of the ratio
${\lD}/{e}$, the drapery function ${\mathcal D}$ is equal to $3/4$ and the
difference $\delta \tilde{I}$ diverges like
\be
\delta \tilde{I} =
{\displaystyle \frac{3 A}{4}} \;
{\displaystyle \frac{e}{\lD}} \;\; .
\label{delta_I_tilde_max}
\ee
Because the energy is measured with a limited accuracy, the relevant quantity
is the averaged value of the difference $\delta \tilde{I}$ which we define as
\be
\delta \tilde{I}_{\Delta E} =
{\displaystyle \frac{1}{\Delta E}} \;
{\displaystyle \int_{E_{S} - \Delta E}^{E_{S}}} \!\! dE \,\,
\delta \tilde{I}(\lD) \;\; .
\ee
This integral is performed on the energy bin that contains the line at
$E_{S} = m_{\chi}$. Close to the line, {\ie} for $E \simeq E_{S}$, the positron
diffusion length $\lD$ is given by expression~(\ref{lD_close_line}) which we
combine with relation~(\ref{delta_I_tilde_max}) to get
\be
\delta \tilde{I}_{\Delta E} =
{\displaystyle \frac{3 A}{2}} \;
\left( {\displaystyle \frac{e}{3.77 \; {\rm kpc}}} \right) \;
{\displaystyle \frac{\epsilon_{S}^{0.65}}{\sqrt{\Delta \epsilon}}}
\;\; .
\label{delta_I_tilde_average}
\ee
For a 100 GeV DM species, this leads to a difference between the fine-grained
and coarse-grained halo integrals of
\be
\delta \tilde{I}_{\Delta E} = 5.8 \times 10^{-3} \;
\sqrt{\displaystyle \frac{E_{S}}{\Delta E}} \;\; .
\ee
A 1\% measurement of the energy translates into a maximal value of the
difference $\delta \tilde{I}_{\Delta E}$ of $0.058 \sim$ 6\%. We therefore
conclude that even in the most extreme situation where the Earth lies in the
region of highest DM density, we find no difference between the results of
Section~\ref{sec:cr_signal} and those derived with the fine-grained
distribution~(\ref{fine_grained}) once the energy is properly averaged.

%
\section{Conclusions}
\label{sec:conclusions}

We have studied whether DM caustics in the halo of the Milky Way can amplify
the flux of cosmic ray antiprotons and positrons which is received at the Earth.
We have used the secondary infall model for our halo which naturally includes
the caustics and assumed that the galactic DM is made of weakly interacting
massive species. We have then taken into account the smearing of the caustics
due to a present-day velocity dispersion of these particles of
$0.03$~cm~s$^{-1}$.

The cosmic ray antiprotons and positrons that are detected at the Earth originate
from a region whose typical size is much larger than the shell thickness or
even the shell separation. The associated horizon probes a large portion of the
Milky Way and the coarse-grained average density~(\ref{DM_shell_QD}) provides
an adequate description. In the solar neighbourhood, the coarse-grained caustic
density is the same as the smooth NFW or isothermal cored distributions usually
assumed in the literature. The difference lies at the centre of the Milky Way.
The lower-envelope density diverges there with an index of $\sim 2.11$, hence
a DM profile steeper than even in the NFW case. The coarse-grained shell density
is assumed to be constant inside a sphere of radius $r_{\rm cut-off}$
whose value is unknown. The smaller this cut-off radius, the more abundant DM at
the galactic centre and consequently the stronger its signal should it reach
the Earth.

We have then computed the antiproton flux which the coarse-grained shell density
yields at the Earth. The reach of the antiproton sphere depends on the cosmic ray
propagation model but is always of order a few kiloparsecs. The MIN configuration
is associated to a rather small antiproton range. The associated signal is not
different from what is derived assuming NFW or isothermal cored distributions.
The MAX propagation model is characterized on the contrary by efficient diffusion
taking over a moderate galactic convection. The antiproton sphere probes the inner
and denser regions of the Milky Way. We find that the antiproton signal is enhanced
by a factor of $\sim 30$ should the conventional smooth NFW profile be replaced by
the coarse-grained shell density~(\ref{DM_shell_QD}) for which a cut-off radius of
300 pc has been assumed. The MED set of propagation parameters corresponds to the
best fit to the B/C data and features the intermediate situation. It leads to the
exciting possibility that the antiproton signal is only boosted above 10 GeV in
the presence of caustics. Depending on the WIMP annihilation cross section, the
antiproton flux could be severely distorted at high energy as shown in
Fig.~\ref{fig:flux_WINO} where no artificial boost factor is required. Thus a
promising window opens up around a few hundreds of Gev where future antiproton
measurements are eagerly awaited.

We are less optimistic for the positron signal. Depending on the cosmic ray
propagation model, the positron flux at the Earth may be enhanced in the
presence of shells with respect to a smooth NFW or isothermal cored DM
profile. However, this situation arises only at low energy where the
observations are already well explained by the sole secondary background
component. Thus DM caustics cannot provide an explanation for the HEAT
excess reported above $\sim$ 10 GeV and consequently produced in the solar
neighbourhood.
The solution so far invoked is based on a WIMP with a hard positron annihilation
spectrum like a Kaluza-Klein particle \citep{Cheng:2002ej} or a neutralino with
a dominant $W^{+}W^{-}$ channel \citep{Hooper:2004bq,Delahaye:2007fr}. Boost
factors of $\sim$ 10 with respect to a smooth NFW DM halo are nevertheless
necessary. Such a value is marginally possible \citep{Lavalle_Maurin}
in scenarios inspired by the
$\Lambda$-CDM N-body numerical simulations which point towards the existence
of numerous and dense DM clumps inside which WIMP annihilation can be enhanced.
We showed that the coarse-grained shell density~(\ref{DM_shell_QD}) does not
provide an alternative to explaining the HEAT excess. However, further data
with higher precision and also a better understanding of the secondary positron
background are needed to firmly exclude caustics as a possible explanation
of the HEAT measurements. It remains very unlikely nevertheless that the present
spectral form could be reproduced by caustics, at least by its coarse-grained
distribution~(\ref{DM_shell_QD}).

We finally explored the possibility that the fine-grained
distribution~(\ref{fine_grained}) could yield a strong positron flux should
the Earth be embedded inside the densest part of a caustic, a region where the DM
density reaches its peak value $\rho_{k,{\sf max}}$. At high energy, the positron
sphere shrinks. Averaging the fine-grained shell density~(\ref{eq:density-caustic})
by its coarse-grained approximation~(\ref{DM_shell_QD}) is no longer possible.
Positrons that are received at the Earth have short diffusion lengths $\lD$ and
hence can be excellent tracers of nearby caustics.
We investigated here the case of a positron line. Positrons that are detected
exactly at the line energy $E = E_{S} \equiv m_{\chi}$ have vanishing diffusion
length and if the Earth sits exactly inside the shell, an enhancement of the
positron signal by a factor of $\sim 13,000$ is naively expected.
However, energy is measured with a limited accuracy and the energy bin of the
line has a non-vanishing width. Averaging $\lD$ over the line bin leads to a
diffusion length which is still far larger than the typical caustic thickness
or even separation. Our study showed no difference between the results of
Section~\ref{sec:cr_signal} and those derived with the fine-grained
distribution~(\ref{fine_grained}) once energy is properly averaged.
We hypothesize that an extremely high energy resolution, presently unavailable,
would allow to detect the nearby caustics.

A word of caution is necessary at this stage though. The analysis of
Section~\ref{sec:draperies} is based on the assumption that cosmic rays
diffuse on the inhomogeneities of the galactic magnetic field. The mean
free path $\lambda_{\rm free}$ of their random walk may be derived from
the space diffusion coefficient $K$ through the canonical relation
\be
K(E) \equiv {\displaystyle \frac{1}{3}} \, \lambda_{\rm free} \, \beta \;\; .
\ee
In this diffusion scheme and with the MED set of parameters, positrons with energy
$\epsilon$ cover on average a distance
\be
\lambda_{\rm free} = 1.1 \times 10^{-4} \; {\rm kpc} \;
\epsilon^{0.7}
\label{lfree_close_line}
\ee
before their next scattering on Alfv\'en waves. Our treatment of cosmic ray
propagation is definitely supported by the fact that $\lambda_{\rm free}$ is much
smaller than the horizon size set by $\lD$ and equation~~(\ref{lD_close_line}).
However, if the nearest DM caustic is very close to the Earth and lies at a distance
which does not exceed $\lambda_{\rm free}$ or if the Earth is embedded inside the
densest part of a shell, the diffusion hypothesis breaks down. Positrons will
essentially drift along the lines of the magnetic field without encountering on
their way any obstacle. Depending on the relative orientation of the local magnetic
field with respect to the Earth and its nearest caustic, we could be possibly exposed
to an intense high-energy positron flux whose evaluation is clearly beyond the scope
of this article.

%
{\small Acknowledgment~: We thank Pierre Brun and Julien Lavalle for
having provided us with a few typical positron and antiproton spectra
arising from WIMP annihilation. RM thanks Sergei Shandarin for
contributions, Niayesh Afshordi, Ed
Bertschinger and Mike Kuhlen for discussions and
LAPTH Annecy for hospitality and 
French programms PNC \& ANR (OTARIE) for travel grants.
Special thanks are due to Mark Vogelsberger for a careful reading of the
manuscript and many useful comments and corrections.
}

%
%
%

\end{document}